\documentclass[preprint,12pt]{elsarticle}

\usepackage{graphicx}
\usepackage{multirow}
\usepackage{amsmath,amssymb,amsfonts}
\usepackage{amsthm}
\usepackage{mathrsfs}
\usepackage[title]{appendix}
\usepackage{textcomp}
\usepackage{manyfoot}
\usepackage{booktabs}
\usepackage{array}
\usepackage{tabularx}
\usepackage{longtable}
\usepackage{ragged2e}
\usepackage{url}
\usepackage{placeins}

\newcolumntype{Y}{>{\RaggedRight\arraybackslash}X}
\newcommand{\botrule}{\bottomrule}
\journal{Biomedical Signal Processing and Control}

\begin{document}

\begin{frontmatter}

\title{State-specific respiratory signatures for affective and stress recognition: Interpretable respiratory markers, autocorrelation lags, and compact CNN models}

\author[inst1]{Andrei Velichko}
\ead{velichkogf@gmail.com}

\author[inst2]{Mehmet Tahir Huyut\corref{cor1}}
\ead{tahir.huyut@erzincan.edu.tr}
\cortext[cor1]{Corresponding author.}

\affiliation[inst1]{organization={Institute of Physics and Technology, Petrozavodsk State University},
    city={Petrozavodsk},
    country={Russia}}

\affiliation[inst2]{organization={Department of Biostatistics and Medical Informatics, Faculty of Medicine, Erzincan Binali Yildirim University},
    city={Erzincan},
    country={Turkey}}

\begin{abstract}
Respiratory activity is a direct and interpretable physiological channel for wearable stress and affective-state recognition, yet many studies emphasize classification accuracy without identifying which respiratory properties separate different states. This work reframes RESP-based recognition as a joint predictive and explanatory problem. Using the chest respiratory channel of the WESAD dataset, we analyze 60 s windows under leave-one-subject-out validation and combine two complementary branches: compact raw-signal one-dimensional convolutional neural networks (1D-CNNs) and physically grouped handcrafted respiratory signatures. The primary application task is binary stress versus non-stress detection, while baseline, stress, amusement, and meditation are additionally analyzed in a one-vs-rest setting to reveal state-specific respiratory markers. The feature space is organized into respiratory timing, breath-to-breath variability, waveform statistics, spectral/time-frequency descriptors, and autocorrelation/nonlinear predictability descriptors, with the raw 60 s signal treated as a sixth representation for the CNN branch. We introduce autocorrelation transition lags (Zpm/Zmp) as interpretable markers of respiratory correlation scale and separately evaluate exploratory forecast-error-growth/Lyapunov-like descriptors. In the final CNN refit setting, the raw-signal model achieved the strongest stress-vs-rest performance (accuracy 96.72\%, macro-F1 95.30\%, MCC 90.61\%). In contrast, compact feature models were stronger for baseline (MCC 65.34\%), amusement (MCC 35.69\%), and especially meditation (MCC 88.65\%). {Strict nested reanalysis of the auxiliary Top-20 stress search yielded mean MCC=82.29\% versus 85.45\% originally, indicating modest selection optimism.} These results show that CNNs are most useful for the practical stress detector, whereas interpretable respiratory signatures provide stronger and more physiologically transparent state-specific markers for several non-stress conditions.
\end{abstract}


\end{frontmatter}

\noindent\textbf{Publication note.} This is the final author archival version of the manuscript. The article was published in \textit{Biomedical Signal Processing and Control}, Volume 129, Part A, 1 January 2027, Article 111373. DOI: \url{https://doi.org/10.1016/j.bspc.2026.111373}.

\section{Introduction}\label{sec:introduction}

Wearable affective computing increasingly requires models that are not only accurate but also compact, subject-independent, and physiologically interpretable. In digital health assistants and mHealth sensor systems, affect and stress recognition can support context-aware feedback, cognitive-resilience monitoring, and timely intervention, but only if the underlying algorithms remain practical for repeated use on wearable or edge-oriented hardware \cite{meigal2023mhealth}. Respiratory activity is a particularly attractive signal for this purpose because breathing is modulated by sympathetic activation, voluntary and involuntary regulation, relaxation, cognitive load, and affective state. Compared with multimodal pipelines, a respiratory-only system has a simpler sensing and deployment path, but it also raises an important question: which properties of the breathing waveform actually distinguish stress, meditation, baseline, and amusement?

The broader wearable-stress literature shows both the promise and the difficulty of this problem. Reviews of biosignal-based stress recognition and wearable affect sensing report rapid growth in machine-learning and deep-learning pipelines, but they also emphasize heterogeneous datasets, small cohorts, sensor-specific biases, and limited cross-subject or real-world generalization \cite{giannakakis2019review,schmidt2019wearable,gedam2021review,bolpagni2024personalized,chatzaki2025overview}. Multimodal systems remain dominant because electrodermal, cardiac, temperature, motion, and respiratory channels capture partially complementary autonomic effects \cite{zhang2024dynamic,abdelfattah2025multimodal}. However, multimodal fusion also increases sensor burden, synchronization complexity, power consumption, and the difficulty of interpreting which physiological pathway drives a decision. This creates a specific gap for single-channel RESP analysis: respiration is repeatedly recognized as informative, but it is less often treated as the sole explanatory object under a strict subject-independent protocol.

{
Recent biomedical machine-learning studies similarly emphasize that limited-data settings require careful control of model capacity and an architecture matched to the available data. Parameter-efficient large-kernel and dual-kernel adapters, efficient transformer designs, and optimization of pre-trained transformer models have been proposed to improve representation learning or reduce overfitting in data-constrained medical applications \cite{zhu2025lka,zhu2026dual,zhu2025bcct,zhu2024optco}. Although these approaches were developed primarily for medical-image classification rather than respiratory time series, they illustrate the broader importance of selecting an appropriate inductive bias and computational complexity when training biomedical models from relatively small datasets.
\par}

From a psychophysiological perspective, breathing is not a passive mechanical signal but a rhythmic behavior coupled to emotion, autonomic regulation, attention, and voluntary control. Earlier work on breathing rhythms and emotions showed that respiratory rate, depth, and timing change with affective state \cite{homma2008breathing}. More recent respiratory-neuroscience and breathwork studies further support the view that breathing rhythm, phase timing, and respiratory regularity modulate neural and mental activity rather than merely reflecting it \cite{ashhad2022breathing,goheen2023lung,balban2023breathwork}. Slow-breathing and breath-control studies suggest that relaxation and meditative states are often accompanied by lower respiratory frequency, altered inhalation/exhalation timing, stronger respiratory sinus arrhythmia, and greater cardiorespiratory coherence \cite{zaccaro2018slowbreathing,noble2019slowdeep,diest2014inhalation,kral2023slower,fincham2023breathwork,birdee2023slowbreathing}. Conversely, stress-related states are frequently associated with faster or less stable breathing and with changes in the spectral distribution of cardiorespiratory oscillations \cite{hernando2016resp_freq,chand2024paced_stress,ritsert2022anxiety,suzuki2023resp_params}. These findings motivate a feature taxonomy that goes beyond mean breathing rate and includes inspiratory timing, respiratory variability, waveform shape, frequency-domain energy, time-frequency structure, and temporal-dependence measures.

{
The temporal scale of respiratory analysis is also physiologically relevant. Respiratory rate can be estimated from short observations, but shortening the observation interval generally reduces repeatability, while breath-to-breath variability and frequency-domain descriptors depend on observing multiple respiratory cycles \cite{nicolo2020rrmonitoring,drummond2020rrprecision,vandenbosch2021breathingvariability}. Accordingly, window length in RESP-based recognition should be viewed not only as a machine-learning hyperparameter, but also as a compromise between physiological feature stability and temporal responsiveness.
\par}

The WESAD dataset is a widely used benchmark for wearable stress and affect recognition. It contains multimodal physiological recordings from 15 subjects during baseline, stress, amusement, and meditation protocols \cite{schmidt2018wesad}. In the original WESAD study, handcrafted physiological features and classical classifiers achieved strong performance, including 80.34\% accuracy for the three-class baseline--stress--amusement problem and 93.12\% accuracy for binary stress versus non-stress recognition \cite{schmidt2018wesad}. Subsequent WESAD work has largely moved toward multimodal fusion, image encodings, deep learning, and personalization \cite{bobade2020stress,ghosh2022stress,pradhan2023hierarchical,kumar2024multimodal,zhang2024dynamic,yang2025image}. These studies are valuable for maximizing recognition accuracy, but they often make it difficult to isolate the standalone contribution of RESP because respiratory effects are entangled with ECG, EDA, EMG, temperature, and motion. At the same time, WESAD sensitivity analyses and wearable affect reviews indicate that respiratory rate and respiratory features are meaningful stress-related signals, although the evidence base for respiration-only WESAD recognition is narrower than for multimodal recognition \cite{iqbal2021sensitivity,schmidt2019wearable,li2023personalized}.

These results make WESAD a useful reference point for analyzing the trade-off between model complexity, subject-independent generalization, and physiological interpretability. In this manuscript, we focus on the chest respiratory (RESP) channel. The primary practical task is binary stress detection, where stress is the positive class and baseline, amusement, and meditation are merged into a non-stress class. This binary formulation is directly relevant for digital health assistants, wearable warning systems, and mobile stress-monitoring tools. However, binary classification alone does not reveal whether different non-stress states have distinct respiratory signatures. Therefore, we additionally analyze the original four WESAD states using one-vs-rest comparisons.

Prior studies on stress and affect recognition have demonstrated that high performance can be obtained with multimodal physiological inputs, image-encoding representations, boosting and deep-learning pipelines, or carefully designed respiratory features \cite{bobade2020stress,ghosh2022stress,kumar2024respboostnet,shan2020resp,barik2023respiration}. Respiratory-only and respiratory-centered studies nevertheless support the feasibility of extracting affective information from breathing patterns: non-contact respiratory stress detection, subject-independent respiration emotion recognition, and respiration-monitoring models for wearable textiles all indicate that a single respiratory stream can contain discriminative temporal structure \cite{shan2020resp,wang2025hierarchical,steinmetzer2025smarttextiles}. However, direct numerical comparison across studies is difficult because label definitions, sensing channels, windowing schemes, and validation protocols often differ. For respiratory-only analysis, recent work by Yang et al. is especially relevant: their physiologically explainable respiratory framework used cycle-level respiratory segmentation, engineered descriptors related to rhythm, depth, and nonlinear dynamics, stacking-based classification, and SHAP-based interpretation for WESAD stress classification \cite{yang2025physio}. This line of work shows that respiration can contain substantial affective information, but it also illustrates that strong results may depend on nontrivial feature design and segmentation heuristics. A complementary question is therefore whether a compact raw-signal model and a transparent feature-screening branch can be evaluated side by side under the same subject-wise protocol.

The deployment motivation is also increasingly supported by the edge-AI and TinyML stress-recognition literature. Lightweight stress models, compact residual networks, and microcontroller-oriented pipelines have been proposed to reduce memory, latency, and energy consumption while preserving useful recognition performance \cite{sim2022edge,gibbs2024tinyml,jaiswal2024tinystressnet,cvetkovic2025bioedgenet,abusamah2025tinyml}. Interpretable and feature-efficient models are similarly attractive because they can reduce computational load and improve clinical transparency \cite{ghose2025lightweight}. These studies do not eliminate the need for physiological interpretation; rather, they strengthen the rationale for comparing a compact raw-signal CNN with a small set of interpretable respiratory features in the same LOSO framework.

Respiratory-only recognition can be approached in two complementary ways. End-to-end neural networks can learn temporal patterns directly from raw one-dimensional windows and may capture information that handcrafted descriptors miss. In contrast, engineered features are easier to interpret because they can be linked to respiratory timing, breath-to-breath variability, amplitude shape, spectral structure, and nonlinear signal dynamics. The present work explicitly organizes the RESP feature space by physical meaning and uses the resulting groups to interpret state-specific behavior. This is important because the practical value of a stress detector is not identical to the scientific value of identifying state-specific respiratory markers: a model may classify stress well while still providing limited insight into why meditation, baseline, or amusement differ.

The current literature also suggests that not all respiratory descriptors have the same level of prior physiological support. Rate, inspiratory/expiratory timing, variability, and spectral/coherence measures are the most established families \cite{diest2014inhalation,hernando2016resp_freq,kral2023slower,soni2019breath_variability}. Respiratory variability has also been linked to anxiety, trauma-related stress, and mental-health indices, which supports the use of breath-to-breath dispersion and cycle-level variability summaries \cite{gazi2021resp_var,ritsert2022anxiety,suzuki2023resp_params}. Temporal-structure and complexity measures, including autocorrelation windows, entropy-like descriptors, and wavelet-based measures, are promising because they summarize regularity, rhythm persistence, and nonstationary variability, but their direct interpretation in stress and meditation recognition is less standardized \cite{oku2022temporal,tiwari2019complexity,goheen2025breathing_dynamics}. This distinction is important for the present study: simple timing and spectral descriptors can be interpreted as relatively direct respiratory markers, whereas Zpm/Zmp and FEG-profile features should be viewed as structured signal descriptors whose physiological meaning must be inferred cautiously.

A central methodological addition of this study is the use of autocorrelation transition lags as interpretable respiratory markers. These descriptors, denoted Zpm and Zmp, measure the lag at which the respiratory autocorrelation changes from positive to negative and from negative to positive, respectively. To our knowledge, these transition-lag descriptors have not been used as explicit markers in prior WESAD RESP stress-recognition pipelines. Unlike more complex forecast-error-growth descriptors, Zpm/Zmp are simple to define and have a direct interpretation as correlation-scale markers of the respiratory waveform. We evaluate these descriptors both at the whole-window level and after breath-cycle aggregation, which allows us to distinguish a typical window-level correlation scale from breath-to-breath variability in that scale.

This study also gives special attention to single-feature and compact-feature analysis. Single respiratory descriptors are not expected to solve the complete recognition problem, but they are useful for discovering interpretable state markers, for checking whether separability is driven by a small number of physiologically meaningful effects, and for identifying cases where respiration alone is weak. Compact feature combinations then test whether a small set of transparent descriptors can approach the performance of raw-signal learning. This design is especially relevant for wearable and digital-health scenarios, where model size, computational cost, and interpretability may be as important as peak accuracy.

The central research question is therefore: which single respiratory markers and compact respiratory feature combinations best distinguish stress, baseline, amusement, and meditation under subject-independent validation, and when is a compact raw-signal CNN preferable to interpretable handcrafted signatures?

The contributions of this work are as follows. First, we restructure RESP-based affect recognition into a primary binary stress-detection task and a secondary one-vs-rest analysis of state-specific respiratory signatures. Second, we propose a physically organized feature taxonomy covering respiratory timing, breath-to-breath variability, waveform statistics, spectral/time-frequency descriptors, and autocorrelation/nonlinear predictability descriptors. Third, we introduce and evaluate Zpm/Zmp autocorrelation transition lags as compact respiratory markers that are rarely considered explicitly in stress-recognition studies. Fourth, we compare interpretable threshold rules and balanced logistic-regression feature combinations with compact 1D-CNN models trained directly on raw RESP windows. Finally, we separately quantify the effect of more complex forecast-error-growth/Lyapunov-like descriptors, treating them as an exploratory extension rather than as the basis of the physiological interpretation. {Quantitatively, the final RESP-only CNN achieved 96.72\% accuracy, 95.30\% macro-F1, and MCC=90.61\% for stress-versus-non-stress detection. Its accuracy is 3.60 percentage points higher than the 93.12\% binary accuracy reported in the original WESAD benchmark \cite{schmidt2018wesad}, while under the present common LOSO protocol it improves MCC by 7.10 points over the best compact interpretable stress model (90.61\% vs.\ 83.51\%). The cross-study accuracy difference is reported as contextual rather than as a strict head-to-head comparison because the input modalities and evaluation settings are not identical.}

The remainder of the paper is organized as follows. Section~\ref{sec:methods} describes the WESAD RESP data, windowing, validation protocol, feature taxonomy, autocorrelation-lag descriptors, compact feature-combination models, and raw-signal CNN. Section~\ref{sec:results} reports single-feature markers, compact feature combinations, the effect of the exploratory FEG-profile descriptors, and the final CNN comparison. Section~\ref{sec:discussion} interprets the state-specific respiratory signatures and limitations of the exploratory feature search. Section~\ref{sec:conclusion} summarizes the main findings.

\section{Materials and methods}\label{sec:methods}

\subsection{Dataset, labels, and analysis tasks}\label{subsec:dataset_tasks}

We used the chest respiratory channel of the WESAD dataset \cite{schmidt2018wesad}. The original protocol labels are baseline, stress, amusement, and meditation. The preprocessing pipeline produced 1984 one-minute RESP windows across 15 subjects; each window contains 3000 samples after downsampling to 50 Hz. Table~\ref{tab:binary_counts} summarizes the state counts and how each state was used in the two analysis settings.

The main application-oriented setting was binary stress detection. In this setting, the stress protocol was treated as the positive class, while baseline, amusement, and meditation were grouped into a non-stress class. This formulation is the most practically relevant one for wearable warning systems and mobile stress-monitoring tools, because the target output is whether the current physiological state is stress-like or not.

In addition, the four original WESAD states were analyzed using a one-vs-rest strategy. For each state, that state was considered positive and all remaining states were considered negative. The one-vs-rest analysis was not introduced as a separate deployment task, but as an interpretive tool for identifying which respiratory feature families are most characteristic of baseline, stress, amusement, and meditation.

\begin{table}[htbp]
\caption{WESAD RESP state counts and their use in the two analysis settings.}\label{tab:binary_counts}%
\centering
\scriptsize
\begin{tabularx}{\textwidth}{@{}p{0.18\textwidth}r p{0.28\textwidth}Y@{}}
\toprule
Original WESAD state & Windows & Role in binary stress task & Role in one-vs-rest signature analysis \\
\midrule
Baseline & 832 & Non-stress class & Positive class in baseline-vs-rest analysis. \\
Stress & 448 & Stress class, main positive label & Positive class in stress-vs-rest analysis; primary application target. \\
Amusement & 226 & Non-stress class & Positive class in amusement-vs-rest analysis. \\
Meditation & 478 & Non-stress class & Positive class in meditation-vs-rest analysis. \\
\midrule
Total & 1984 & Stress vs non-stress & Four separate one-vs-rest comparisons. \\
\botrule
\end{tabularx}
\end{table}

\begin{figure}[htbp]
\centering
\begin{minipage}{0.32\textwidth}
\centering
\includegraphics[width=\linewidth]{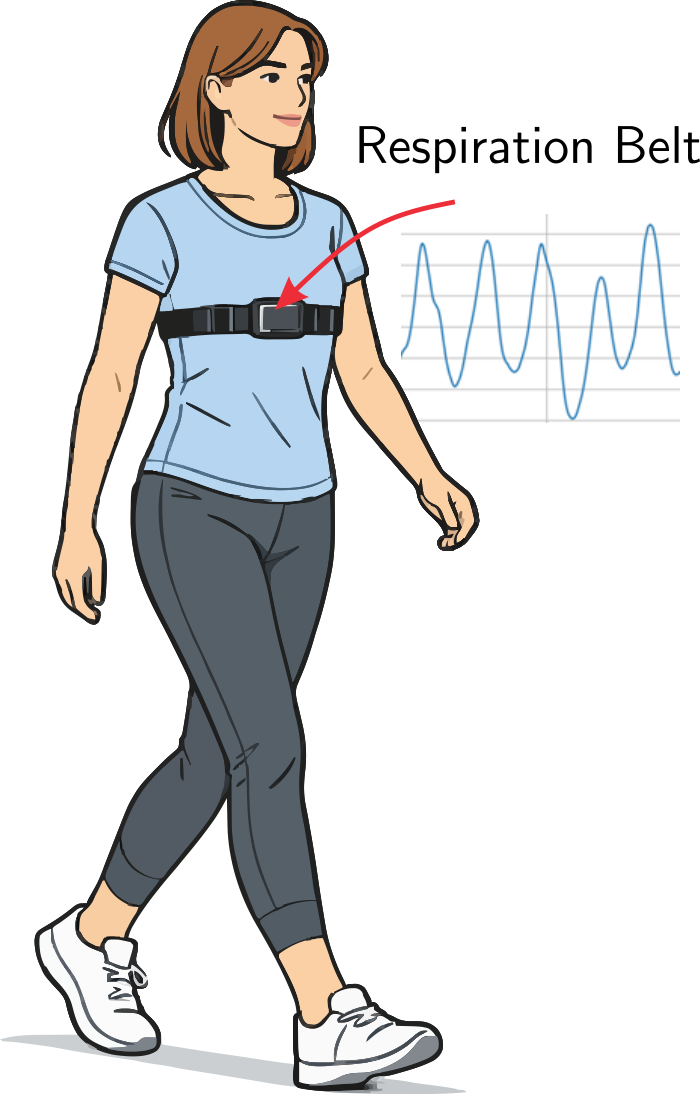}\\
\small (a) Chest respiration belt.
\end{minipage}\hfill
\begin{minipage}{0.64\textwidth}
\centering
\includegraphics[width=\linewidth]{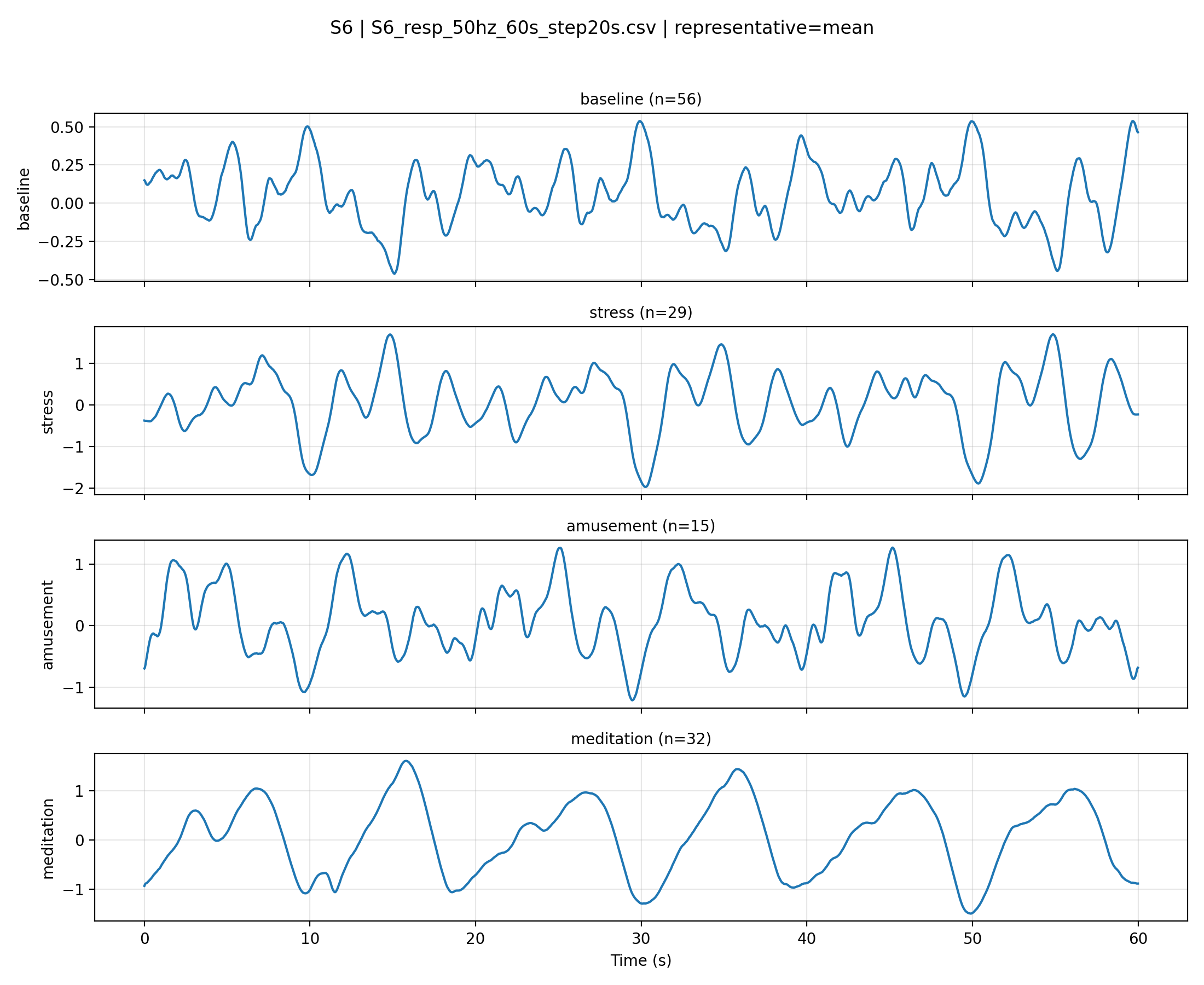}\\
\small (b) Representative one-minute RESP series.
\end{minipage}
\caption{Respiratory data used in the study. The primary binary analysis uses stress as the positive class and groups baseline, amusement, and meditation as non-stress. The secondary one-vs-rest analysis is used to identify state-specific respiratory signatures.}
\label{fig:data_overview}
\end{figure}

The original acquisition setup and representative respiratory series are shown in Fig.~\ref{fig:data_overview}.

\subsection{Windowing, preprocessing, and validation}\label{subsec:preprocessing}

The raw chest RESP signal was segmented using a common windowing procedure for all models and feature sets. A 10 s margin was excluded near label boundaries to reduce transition contamination. The remaining signal was divided into 60 s windows with a 20 s step. A fourth-order Butterworth low-pass filter was applied, the signal was downsampled to 50 Hz, and each window was normalized. This produced fixed-length one-dimensional inputs of length $N=3000$ for the CNN and a common window basis for feature extraction.

{
The 60 s duration was selected as a physiologically motivated compromise rather than as a strict minimum requirement. In adults, spontaneous respiratory rates commonly fall in a range that yields roughly a dozen or more breaths per minute, and one-minute observation remains a conventional reference interval for respiratory-rate assessment \cite{nicolo2020rrmonitoring}. Shorter observation periods can provide faster rate estimates, but their repeatability decreases as the interval is shortened, whereas respiratory-variability measures require enough cycles to characterize breath-to-breath fluctuations \cite{drummond2020rrprecision,vandenbosch2021breathingvariability}. In the present data, the median number of detected cycles per 60 s window was 13.0 [9.0--16.0], supporting the use of this duration for the timing, variability, spectral, and autocorrelation-derived descriptors analyzed here. The 20 s stride provides a refreshed model output every 20 s while preserving the 60 s physiological context; it should not be interpreted as a fixed 20 s state-transition latency.
\par}

All reported models use subject-independent leave-one-subject-out (LOSO) validation. In each fold, one subject is held out for testing and the remaining subjects are used for training, feature normalization, threshold selection, and model fitting. Thus, all thresholds, imputers, scalers, and classifier parameters are learned without access to the held-out subject.

Respiratory-cycle detection was also subjected to internal quality control because several timing and breath-cycle aggregation features depend on reliable cycle segmentation. The automatic trough--peak--trough detector was evaluated on all 1,984 60 s RESP windows. A window was considered valid when at least three complete respiratory cycles were detected without fallback estimation. Valid cycle detection was obtained in 1,984/1,984 windows (100.00\%), and fallback estimation was not required in any window. Across all windows, the median number of detected cycles was 13.0 [9.0--16.0], corresponding to a mean cycle rate of 12.80 $\pm$ 4.42 cycles/min. The mean cycle coverage was 0.996 $\pm$ 0.135, and the mean dominant respiratory frequency was 0.217 $\pm$ 0.079 Hz. These values indicate that the automatic detector provided complete internal coverage of the analyzed WESAD RESP windows, although the dataset does not include manual breath-by-breath annotations for external boundary validation.

\subsection{Overall study workflow and block diagram}\label{subsec:workflow}

\begin{figure}[!htbp]
\centering
\includegraphics[width=0.94\textwidth]{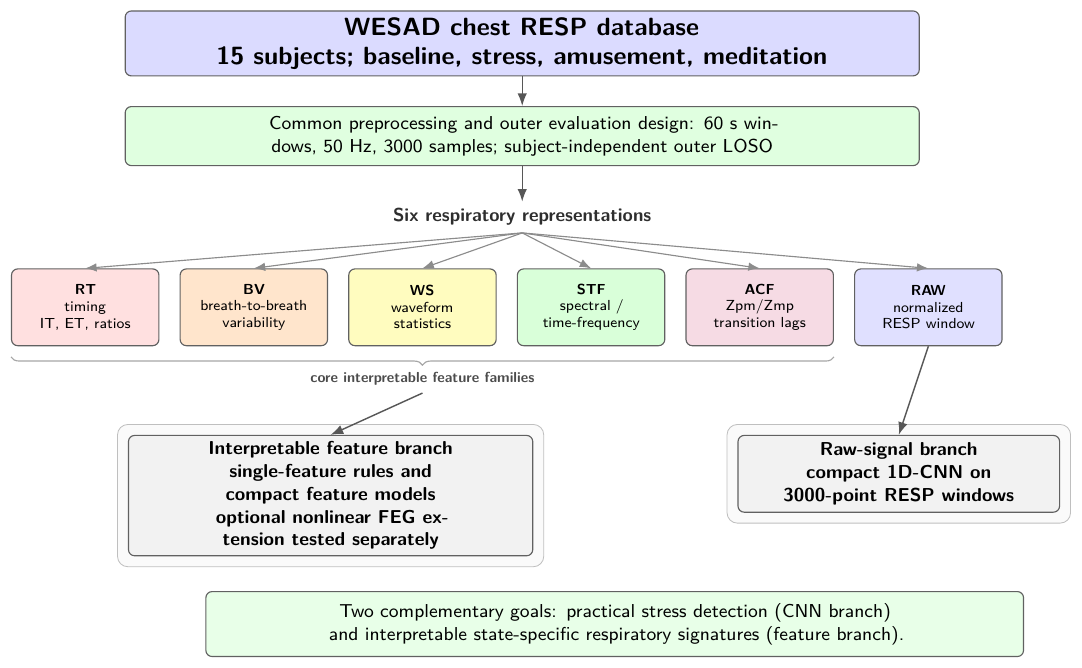}
\caption{{Overall workflow and respiratory representations. The WESAD chest RESP channel is converted into normalized 60 s windows and represented through five core interpretable feature families plus the raw-signal representation. The feature branch is used for single-feature marker screening and compact interpretable models, whereas the raw branch is used by the compact 1D-CNN. Detailed outer-LOSO training/testing paths for the two branches are shown separately in Figs.~\ref{fig:feature_loso_workflow} and~\ref{fig:cnn_loso_workflow}.}}
\label{fig:workflow}
\end{figure}

The full workflow is shown in Fig.~\ref{fig:workflow}. The analysis starts from the WESAD chest RESP database and label-aware 60 s windowing. Each normalized window is represented by six complementary descriptions. Five descriptions are physically interpretable: respiratory timing (RT), breath-to-breath variability (BV), waveform statistics (WS), spectral/time-frequency descriptors (STF), and autocorrelation transition-lag descriptors (ACF; Zpm/Zmp). These groups form the core interpretable feature branch. The sixth description is the raw normalized RESP window (RAW), which enters the compact 1D-CNN branch.

{
To make the separation between training and testing explicit, the original workflow presentation has been expanded into one overview diagram and two branch-specific outer-LOSO diagrams. Figure~\ref{fig:workflow} summarizes the shared preprocessing, respiratory representations, and the two analysis branches. Figure~\ref{fig:feature_loso_workflow} details the feature-model outer-LOSO path, in which fold-specific thresholds, scaling statistics, lookup tables, and classifier parameters are fitted from the 14 non-held-out subjects before evaluation on the held-out subject. Figure~\ref{fig:cnn_loso_workflow} details the CNN path, including its internal TRAIN/VAL split for training-time selection and the subsequent refit on the complete 14-subject outer-training pool. The exploratory feature-combination search itself remains distinct from this fold-wise fitting path and is described separately in Section~\ref{subsec:feature_models}.
\par}

The interpretable branch uses single-feature threshold rules for marker screening and compact models based on binary Laplace threshold patterns or balanced logistic regression. The raw-signal branch trains a compact 1D-CNN on 3000-point windows. All thresholds, scalers, imputers, model parameters, and CNN training procedures are fitted inside the training subjects of each LOSO fold. The workflow separates the core interpretable feature families from the optional nonlinear FEG-Pro extension, so that physiologically transparent signatures can be evaluated before adding more complex predictability descriptors.

\subsection{Physical grouping of respiratory representations}\label{subsec:feature_groups}

The RESP window was described using the six representation groups listed in Table~\ref{tab:feature_families}. The first five groups form the interpretable feature branch; the raw signal forms the CNN branch. Features adopted from the physiologically explainable respiratory-feature framework of Yang et al.~\cite{yang2025physio} are explicitly marked in the source column. Features marked as ``this study'' are newly introduced or newly adapted in the present WESAD RESP pipeline. This distinction is important because the main physiological conclusions are based on the core interpretable groups, whereas the more complex forecast-error-growth variables are treated as an exploratory nonlinear extension. In the workflow diagram, the autocorrelation transition-lag component is labelled ACF, while the broader table abbreviation ANP is retained for the combined autocorrelation/nonlinear predictability family used in the result tables.

\begin{table}[htbp]
\caption{Respiratory representation groups used in the study.}\label{tab:feature_families}%
\centering
\scriptsize
\begin{tabularx}{\textwidth}{@{}p{0.18\textwidth}p{0.07\textwidth}p{0.23\textwidth}p{0.18\textwidth}Y@{}}
\toprule
Group & Abbrev. & Representative variables & Source & Physical interpretation \\
\midrule
Respiratory timing and cycle morphology & RT & BR, IT, ET, IT\_ratio, D, RSBI, SD1, SD2 & Yang et al.~\cite{yang2025physio} & Timing of the respiratory cycle, inspiratory/expiratory phase organization, depth, and cycle-interval geometry. \\
Breath-to-breath variability & BV & BR\_cv, IT\_cv, ET\_cv, IT\_ratio\_cv, D\_cv, RSBI\_cv & Yang et al.~\cite{yang2025physio} & Stability or instability of respiratory timing and depth across consecutive cycles. \\
Waveform statistics & WS & Avg, Std, Rat, Skw, Kur, min, max, quantiles, IQR, RMS, energy, slope, zero-crossing rate & Yang et al.~\cite{yang2025physio} plus standard window statistics & Amplitude distribution, waveform asymmetry, dispersion, total energy, trend, and sign-change behavior. \\
Spectral and time-frequency descriptors & STF & Fre, Fre1--Fre6, FFT band powers, Wer, Wee, We, Wse & Yang et al.~\cite{yang2025physio} plus FFT bands added in this study & Dominant breathing frequency, redistribution of power over frequency bands, and wavelet energy/entropy structure. \\
Autocorrelation and nonlinear predictability & ANP & Zpm, Zmp, BC-median Zmp, BC-CV Zmp, selected FEG-profile descriptors & Zpm/Zmp and FEG-Pro extension introduced in this study & Correlation-scale structure, breath-cycle variability of autocorrelation lags, and exploratory nonlinear predictability. \\
Raw RESP signal & RAW & 3000-point normalized 60 s window & WESAD RESP channel~\cite{schmidt2018wesad} & Direct waveform representation used by the compact 1D-CNN. \\
\botrule
\end{tabularx}
\end{table}

\subsubsection{Respiratory timing and variability features}\label{subsubsec:timing_variability}

Respiratory timing and variability descriptors follow the physiologically meaningful respiratory feature family used by Yang et al.~\cite{yang2025physio}, with implementation adapted to the present 60 s WESAD RESP windows. Peak-to-valley respiratory cycles were detected after smoothing and low-pass preprocessing. Inspiration time (IT) was defined as the duration from trough to subsequent peak, expiration time (ET) as the duration from peak to subsequent trough, and breath-to-breath interval (BB) as the time between successive comparable cycle landmarks. The breathing rate for a cycle was approximated as $BR_i=60/BB_i$, and the inspiratory fraction was defined as $IT\_ratio_i = IT_i/(IT_i+ET_i)$.

Respiratory depth $D_i$ was estimated from peak-to-trough amplitude, and the rapid shallow breathing index was approximated as $RSBI_i=BR_i/D_i$. For any cycle-level descriptor $q_i$, breath-to-breath variability was summarized by $q\_cv = \operatorname{std}(q_i)/(\lvert\operatorname{mean}(q_i)\rvert+\epsilon)$, where $\epsilon$ prevents division by zero. Poincare descriptors SD1 and SD2 summarize short- and longer-scale variability of breath-to-breath intervals. Approximate entropy (Appro) and sample entropy (Sample) were also computed from the smoothed respiratory window as irregularity descriptors. The timing and variability variables used in the manuscript are listed in Table~\ref{tab:timing_features}.

\begin{table}[htbp]
\caption{Respiratory timing, variability, and entropy descriptors.}\label{tab:timing_features}%
\centering
\scriptsize
\begin{tabularx}{\textwidth}{@{}p{0.14\textwidth}p{0.09\textwidth}p{0.18\textwidth}Y p{0.16\textwidth}@{}}
\toprule
Feature & Group & Level & Calculation / meaning & Source \\
\midrule
BR & RT & Cycle aggregated & Mean breathing rate, $60/BB_i$, where $BB_i$ is the breath-to-breath interval. & Yang et al.~\cite{yang2025physio} \\
IT, ET & RT & Cycle aggregated & Mean inspiratory and expiratory durations. & Yang et al.~\cite{yang2025physio} \\
IT\_ratio & RT & Cycle aggregated & Fraction of cycle time occupied by inspiration, $IT/(IT+ET)$. & Yang et al.~\cite{yang2025physio} \\
D & RT & Cycle aggregated & Peak-to-trough respiratory depth. & Yang et al.~\cite{yang2025physio} \\
RSBI & RT & Cycle aggregated & Rapid shallow breathing index, approximated as $BR/D$. & Yang et al.~\cite{yang2025physio} \\
BR\_cv, IT\_cv, ET\_cv & BV & Cycle aggregated & Relative variability of rate, inspiratory time, and expiratory time. & Yang et al.~\cite{yang2025physio} \\
IT\_ratio\_cv & BV & Cycle aggregated & Relative variability of inspiratory fraction; one of the main stress-signature candidates. & Yang et al.~\cite{yang2025physio} \\
D\_cv, RSBI\_cv & BV & Cycle aggregated & Relative variability of depth and shallow-breathing index. & Yang et al.~\cite{yang2025physio} \\
SD1, SD2 & BV & Cycle aggregated & Poincare descriptors of short- and longer-scale breath-interval variability. & Yang et al.~\cite{yang2025physio} \\
Appro, Sample & BV/ANP & Whole window & Approximate entropy and sample entropy; irregularity of the respiratory series. & Yang et al.~\cite{yang2025physio} \\
\botrule
\end{tabularx}
\end{table}

\subsubsection{Waveform statistics}\label{subsubsec:waveform_statistics}

Waveform-statistical features describe the amplitude distribution and shape of the normalized 60 s respiratory window. The Yang general time-domain variables Avg, Std, Rat, Skw, and Kur were included following Yang et al.~\cite{yang2025physio}. Here Avg is the mean window amplitude, Std is the standard deviation, Rat is the ratio between the maximum value and the mean value, Skw is skewness, and Kur is non-excess kurtosis. We additionally used standard whole-window statistics: minimum, maximum, median, quartiles, interquartile range, mean absolute value, root-mean-square amplitude, total energy, linear slope, and zero-crossing rate. These features capture waveform amplitude, dispersion, asymmetry, and the presence of slow trends or rapid sign changes. Table~\ref{tab:waveform_features} summarizes the waveform-statistical group.

\begin{table}[htbp]
\caption{Waveform-statistical descriptors of the 60 s respiratory segment.}\label{tab:waveform_features}%
\centering
\scriptsize
\begin{tabularx}{\textwidth}{@{}p{0.16\textwidth}p{0.12\textwidth}Y p{0.18\textwidth}@{}}
\toprule
Feature & Level & Calculation / interpretation & Source \\
\midrule
Avg, Std & WW & Mean and standard deviation of the RESP window. & Yang et al.~\cite{yang2025physio} \\
Rat & WW & Ratio of maximum value to mean value; implemented with a safe denominator. & Yang et al.~\cite{yang2025physio} \\
Skw, Kur & WW & Skewness and non-excess kurtosis; waveform asymmetry and tail behavior. & Yang et al.~\cite{yang2025physio} \\
min, max, median & WW & Basic amplitude range and central tendency. & Standard window statistics \\
q25, q75, IQR & WW & Robust amplitude distribution summaries. & Standard window statistics \\
RMS, abs\_mean, energy & WW & Root-mean-square amplitude, mean absolute amplitude, and total signal energy. & Standard window statistics \\
slope & WW & Linear trend across the 60 s window. & Standard window statistics \\
zero-crossing rate & WW & Number of sign changes per sample/time; frequency-like waveform marker. & Standard window statistics \\
\botrule
\end{tabularx}
\end{table}

\subsubsection{Spectral, frequency-band, and wavelet descriptors}\label{subsubsec:spectral_wavelet}

Spectral features quantify where respiratory energy is located in frequency. Two complementary spectral representations were used. First, five broad FFT band-power features were computed from the demeaned respiratory window. For a window $x(t)$, the discrete Fourier spectrum was estimated and the power was summed inside predefined frequency intervals. The resulting FFT band powers were 0.05--0.20, 0.20--0.50, 0.50--1.00, 1.00--2.00, and 2.00--5.00 Hz. These bands were added in the present study to provide a transparent frequency-energy representation from slow respiratory oscillations to higher waveform components.

Second, Yang-style frequency descriptors were computed following the general respiratory feature set of Yang et al.~\cite{yang2025physio}. The variable Fre is the dominant spectral frequency in the 0.05--1.00 Hz search band. Fre1--Fre6 are power sums in ordered frequency bands: Fre1, 0--0.10 Hz; Fre2, 0.10--0.20 Hz; Fre3, 0.20--0.30 Hz; Fre4, 0.30--0.40 Hz; Fre5, 0.40--0.70 Hz; and Fre6, 0.70--1.00 Hz. Thus, Fre6 represents high-frequency respiratory spectral energy and was later evaluated as a candidate stress marker.

Time-frequency structure was described using wavelet descriptors from the Yang general feature set. Wer is a wavelet energy ratio, Wee is wavelet energy entropy across subbands, We is a wavelet entropy descriptor, and Wse is wavelet singular entropy. These variables summarize whether respiratory energy is concentrated in a small number of wavelet components or distributed across time-frequency scales. The complete STF group is listed in Table~\ref{tab:spectral_features}.

\begin{table}[htbp]
\caption{Frequency-band and wavelet descriptors used in the spectral/time-frequency group.}\label{tab:spectral_features}%
\centering
\scriptsize
\begin{tabularx}{\textwidth}{@{}p{0.15\textwidth}p{0.20\textwidth}Y p{0.18\textwidth}@{}}
\toprule
Feature & Frequency range / level & Calculation / interpretation & Source \\
\midrule
FFT$_{0.05-0.20}$ & 0.05--0.20 Hz & Power in slow respiratory oscillations. & This study \\
FFT$_{0.20-0.50}$ & 0.20--0.50 Hz & Power in the common adult breathing-rate range and nearby components. & This study \\
FFT$_{0.50-1.00}$ & 0.50--1.00 Hz & Power in faster respiratory components and harmonics. & This study \\
FFT$_{1.00-2.00}$ & 1.00--2.00 Hz & Higher-frequency respiratory/waveform components. & This study \\
FFT$_{2.00-5.00}$ & 2.00--5.00 Hz & High-frequency components retained after preprocessing. & This study \\
Fre & 0.05--1.00 Hz & Dominant/main frequency, selected as the peak spectral frequency in this search band. & Yang et al.~\cite{yang2025physio} \\
Fre1 & 0--0.10 Hz & Spectral power sum in the lowest Yang frequency band. & Yang et al.~\cite{yang2025physio} \\
Fre2 & 0.10--0.20 Hz & Spectral power sum in the second Yang frequency band. & Yang et al.~\cite{yang2025physio} \\
Fre3 & 0.20--0.30 Hz & Spectral power sum in the third Yang frequency band. & Yang et al.~\cite{yang2025physio} \\
Fre4 & 0.30--0.40 Hz & Spectral power sum in the fourth Yang frequency band. & Yang et al.~\cite{yang2025physio} \\
Fre5 & 0.40--0.70 Hz & Mid/high-frequency respiratory spectral power. & Yang et al.~\cite{yang2025physio} \\
Fre6 & 0.70--1.00 Hz & High-frequency respiratory spectral power. & Yang et al.~\cite{yang2025physio} \\
Wer & wavelet subbands & Wavelet energy ratio of the first detail subband. & Yang et al.~\cite{yang2025physio} \\
Wee & wavelet subbands & Wavelet energy entropy across subbands. & Yang et al.~\cite{yang2025physio} \\
We, Wse & wavelet subbands & Wavelet entropy and wavelet singular entropy. & Yang et al.~\cite{yang2025physio} \\
\botrule
\end{tabularx}
\end{table}

\subsection{Autocorrelation transition lags}\label{subsec:zpm_zmp}

Autocorrelation transition lags were introduced in this study as new interpretable markers of respiratory correlation structure. For a windowed respiratory signal $x_t$, the normalized autocorrelation $R(\tau)$ was computed for increasing lags $\tau$. Zpm was defined as the first lag at which $R(\tau)$ changes from positive to negative. Zmp was defined as the subsequent lag at which $R(\tau)$ changes from negative to positive. In qualitative terms, Zpm measures when the waveform first loses positive self-similarity, whereas Zmp measures when the autocorrelation returns from the negative phase to a positive phase. These lags therefore reflect the dominant temporal scale, regularity, and alternation geometry of the breathing waveform rather than simply the breathing rate.

The Zpm/Zmp family was evaluated at several aggregation levels. Whole-window Zpm/Zmp were computed once from the complete 60 s segment. Breath-cycle versions were computed over local respiratory-cycle segments and then summarized across cycles. The most interpretable aggregated variables are BC-median Zmp, which reflects the typical autocorrelation transition scale of individual breaths, and BC-CV Zmp, which reflects breath-to-breath instability of that transition scale. Table~\ref{tab:zpm_zmp_features} lists the autocorrelation-lag descriptors introduced in this study.

\begin{table}[htbp]
\caption{Autocorrelation transition-lag descriptors introduced in this study.}\label{tab:zpm_zmp_features}%
\centering
\scriptsize
\begin{tabularx}{\textwidth}{@{}p{0.18\textwidth}p{0.16\textwidth}Y p{0.16\textwidth}@{}}
\toprule
Feature notation & Level & Meaning & Source \\
\midrule
Zpm & WW or BC & First positive-to-negative transition lag of the normalized autocorrelation function. & This study \\
Zmp & WW or BC & Subsequent negative-to-positive transition lag of the normalized autocorrelation function. & This study \\
BC-median Zmp & Breath-cycle aggregation & Median Zmp across local breath-cycle segments; typical cycle-level correlation scale. & This study \\
BC-CV Zmp & Breath-cycle aggregation & Relative breath-to-breath variability of Zmp; instability of correlation scale. & This study \\
BC-IQR / BC-range Zpm/Zmp & Breath-cycle aggregation & Robust or full spread of autocorrelation transition lags across cycles. & This study \\
SEG3 Zpm/Zmp & Three-segment aggregation & Variation of autocorrelation transition lags across three 20 s subwindows. & This study \\
\botrule
\end{tabularx}
\end{table}

\subsection{Feature aggregation levels and notation}\label{subsec:aggregation}

Several descriptors were computed at more than one aggregation level because a respiratory state may affect either the average value of a descriptor or its instability over time. Whole-window features (WW) were calculated once from the full 60 s segment. Breath-cycle features (BC) were calculated over detected cycles or local cycle-aligned segments and then aggregated within the window. Three-segment features (SEG3) were calculated after dividing the 60 s window into three equal 20 s subwindows. The main text emphasizes WW and BC descriptors because they are easier to interpret physiologically; SEG3 descriptors are treated as secondary exploratory variables. Table~\ref{tab:aggregation} defines the aggregation notation used in feature names and result tables.

\begin{table}[htbp]
\caption{Aggregation notation for respiratory features.}\label{tab:aggregation}%
\centering
\scriptsize
\begin{tabularx}{\textwidth}{@{}p{0.17\textwidth}p{0.22\textwidth}Y@{}}
\toprule
Notation & Example & Meaning \\
\midrule
WW & WW Fre6, WW skewness & Descriptor computed once over the complete 60 s window. \\
BC-mean & BC-mean Zmp & Mean value of a descriptor across detected respiratory cycles. \\
BC-median & BC-median Zmp & Median value across detected respiratory cycles. \\
BC-CV & BC-CV Zmp & Relative variability across cycles, computed as standard deviation divided by absolute mean. \\
BC-IQR / BC-range & BC-IQR Zmp & Robust or full range of cycle-level descriptor values. \\
BC-slope / BC-diff & BC-slope Zmp & Linear trend or last-minus-first difference across cycles. \\
SEG3-mean / SEG3-CV & SEG3-mean Zmp & Descriptor aggregated across three equal 20 s subwindows. \\
\botrule
\end{tabularx}
\end{table}

\subsection{Exploratory forecast-error-growth descriptors}\label{subsec:fegpro}

In addition to the core interpretable respiratory features, we evaluated an exploratory nonlinear extension based on forecast-error-growth profiling (FEG-Pro) for scalar time series \cite{velichko2026fegpro}. The same notation is used here and in Appendix~\ref{app:feg_defs}. For a respiratory segment, nearest-neighbor multi-horizon forecasts produce a geometrically averaged forecast error $G(\tau)$ and a log error-growth profile
\begin{equation}
    L(\tau)=\log G(\tau),
\end{equation}
where $\tau$ is the forecast horizon. Features in this group summarize the shape, smoothness, and distributional uncertainty of $L(\tau)$ rather than direct clinical respiratory quantities.

The main global descriptor is the one-line FEG slope $\lambda_{\mathrm{FEG}}$, obtained from $L(\tau) \approx a+\lambda_{\mathrm{FEG}}\tau$; the corresponding unsigned descriptor is $|\lambda_{\mathrm{FEG}}|$. To represent non-uniform profile geometry, we also used early and late two-line slopes $b_L$ and $b_R$, a quadratic mid-horizon slope $b_Q+2c_Q\tau_{\mathrm{mid}}$, $R^2$-type profile-fit descriptors, piecewise/curvature improvement measures, residual-shape descriptors derived from $r(\tau)=L(\tau)-\widehat{L}_Q(\tau)$, monotonicity fractions, slope-stability scores, and forecast-error distribution entropy (FEDE) descriptors $H_{\mathrm{FEDE}}$. Whole-window descriptors are used directly. Breath-cycle descriptors are prefixed by the aggregation notation in Table~\ref{tab:aggregation}; for example, BC-CV $b_L$ denotes the relative breath-to-breath variability of the early local FEG slope, whereas BC-median $\Delta H_{\mathrm{FEDE}}$ denotes a typical cycle-level entropy-change descriptor.

These variables are treated as optional nonlinear predictability and variability summaries. The main physiological interpretation is therefore based first on the core interpretable set, excluding FEG-Pro variables. The full nonlinear extension is then evaluated separately to test whether these more complex descriptors improve compact threshold or balanced-logistic-regression models. The computational outline is given in Appendix~\ref{app:feg_defs}.

\begin{table}[!htbp]
\caption{Exploratory FEG-Pro descriptors considered as nonlinear predictability markers. The notation matches Appendix~\ref{app:feg_defs}.}\label{tab:feg_descriptors}%
\centering
\scriptsize
\begin{tabularx}{\textwidth}{@{}p{0.25\textwidth}p{0.23\textwidth}Y@{}}
\toprule
Notation in this paper & Descriptor family & Qualitative interpretation \\
\midrule
$\lambda_{\mathrm{FEG}}$, $|\lambda_{\mathrm{FEG}}|$ & Global growth & Signed and absolute slope of the log forecast-error-growth profile $L(\tau)$. \\
$b_L$, $b_R$ & Two-line local slopes & Early- and late-horizon error-growth tendencies of $L(\tau)$. \\
$b_Q+2c_Q\tau_{\mathrm{mid}}$, $c_Q$ & Quadratic shape & Mid-horizon slope and curvature of a quadratic approximation to $L(\tau)$. \\
$R^2_{\mathrm{single}}$, $R^2_{\mathrm{two}}$, $R^2_{\mathrm{quad}}$ & Profile fit quality & How well one-line, two-line, and quadratic models describe $L(\tau)$. \\
Two-line improvement; curvature gain & Shape improvement & Added explanatory value of piecewise or curved profile representations. \\
$\operatorname{SD}(r)$, residual roughness, residual second difference & Residual-shape variability & Irregularity left after subtracting the smooth fitted profile $\widehat{L}_Q(\tau)$. \\
Monotonicity fractions & Directional consistency & Fraction of adjacent horizon steps for which $L(\tau)$ increases or decreases. \\
Slope-stability and reliability scores & Reliability/stability & Consistency of estimated growth tendencies across fitted profile regions. \\
$H_{\mathrm{FEDE}}$, $\Delta H_{\mathrm{FEDE}}$, FEDE early slope & Forecast-error distribution entropy & Entropy, early/late change, and early-horizon slope of normalized forecast-error-difference distributions. \\
FEDE h1, hlast, min, max & FEDE horizon summaries & Normalized early, late, minimum, and maximum entropy/difference levels. \\
\botrule
\end{tabularx}
\end{table}
\FloatBarrier

\subsection{Single-feature threshold screening}\label{subsec:single_thresholds}

Single-feature screening was performed for each binary task using training-fold thresholds. For a given feature $f$, a threshold $\theta$ and direction were selected using only the training subjects of a LOSO fold. The positive prediction was then defined as either $f\geq\theta$ or $f\leq\theta$, depending on which direction maximized the selected training criterion. The main screening criterion was MCC; positive-class F1 was used in additional robustness checks for rare or recall-sensitive cases. Candidate thresholds were generated from the training-fold feature values. If the number of unique training values exceeded 301, a quantile grid of 301 candidate thresholds was used; otherwise, midpoints between sorted unique values were tested, together with extreme thresholds allowing all-negative or all-positive predictions. Thus, each threshold, its direction, and any imputation statistic were learned only from the training subjects of the corresponding LOSO fold.

This procedure yields one fold-specific threshold per feature and held-out subject. For reporting, fold results were pooled or averaged across the 15 LOSO folds. These single-feature thresholds also served as the basis for binary threshold-pattern models.

\subsection{Compact feature-combination models}\label{subsec:feature_models}

Two compact interpretable modeling strategies were considered. First, binary threshold-pattern models used the fold-specific single-feature thresholds to convert each feature into a binary indicator. A combination of $k$ features therefore becomes a binary code of length $k$. For each LOSO fold, the binary code probabilities were estimated from the training subjects only using a Laplace-stabilized lookup table. If $c$ denotes a binary code, the positive-class score was defined as
\[
\widehat{s}(c)=\frac{n_+(c)+\alpha\pi}{n(c)+\alpha},
\]
where $n_+(c)$ is the number of positive training samples with code $c$, $n(c)$ is the total number of training samples with code $c$, $\pi$ is the positive-class prior in the corresponding training fold, and $\alpha=1.0$ was fixed a priori. A score threshold was then selected on the training fold and applied unchanged to the held-out subject. This model is referred to as the binary Laplace threshold-pattern model (BL) in the result tables. No feature threshold, binary code statistic, lookup table, or score threshold was fitted on the held-out subject.

For one-vs-rest state-signature analysis, the threshold-pattern combinations used the strongest single-feature threshold markers as candidate inputs. Up to 50 candidate features per target state were retained after requiring single-feature MCC $\geq 0.20$. Size-1 results were taken directly from the precomputed single-feature threshold-MCC table. Combinations of size 2--5 were evaluated with the same Laplace-smoothed lookup strategy. To keep the exploratory search computationally bounded, no more than 500,000 combinations were evaluated for any given size. When the full search space exceeded this limit, candidate combinations were generated by extending the best lower-order patterns and completing the remaining quota with reproducible random combinations; the random seed was fixed at 1777. This design was used to find compact threshold signatures and was not a nested feature-selection validation.

Second, balanced logistic regression was trained on continuous feature values. Within each LOSO fold, missing values were imputed using training-fold medians, features were standardized using training-fold means and standard deviations, and logistic regression was fitted with class-weight balancing. Balanced logistic regression provides a compact continuous-feature classifier and was used as the main non-neural model for stress detection.

For the auxiliary continuous-feature multi-classifier search used to check compact feature subsets, combinations of size 2 and 3 were evaluated exhaustively over the full 52-feature core pool. For larger subsets, an exploratory top-ranked search was used: combinations of size 4--7 were evaluated over the top 20 single features ranked by LOSO MCC. This resulted in 1,326 two-feature, 22,100 three-feature, 4,845 four-feature, 15,504 five-feature, 38,760 six-feature, and 77,520 seven-feature subsets, for a total of 160,055 tested combinations. For each combination, lightweight classifiers were evaluated under LOSO validation, including balanced logistic regression, LDA, Gaussian Naive Bayes, a depth-2 balanced decision tree, and distance-weighted kNN. The random seed was fixed at 42 for reproducibility. This search was exploratory and was not a nested feature-selection procedure; it is intended to identify compact signatures and feature families rather than to claim an independent feature-selection validation.

{
To directly quantify possible selection optimism in this Top-20 procedure, an additional strict nested subject-independent sensitivity analysis was performed for the binary stress task. In each outer LOSO fold, the held-out subject was excluded before any feature ranking or model selection. The remaining 14 subjects formed the outer-training pool and were evaluated by an inner five-fold GroupKFold split by subject. All 52 core features were re-ranked within that outer-training pool using single-feature balanced logistic regression, and a fold-specific Top-20 candidate set was constructed. Within the same inner validation, four-feature subsets were evaluated exhaustively (4,845 combinations), whereas sizes 5--7 were evaluated by deterministic beam expansion from the best lower-order candidates (beam width 300) to keep the nested calculation computationally tractable. Balanced logistic regression was fixed as the confirmatory model. The selected subset was then refitted using all 14 outer-training subjects and applied once to the held-out subject. Thus, the outer test subject contributed neither to feature ranking, Top-20 construction, subset selection, preprocessing statistics, nor model fitting. This nested calculation is reported as a sensitivity analysis of the auxiliary Top-20 search rather than as a replacement for the broader exploratory state-signature analysis.
\par}

\begin{figure}[!htbp]
\centering
\includegraphics[width=0.92\textwidth]{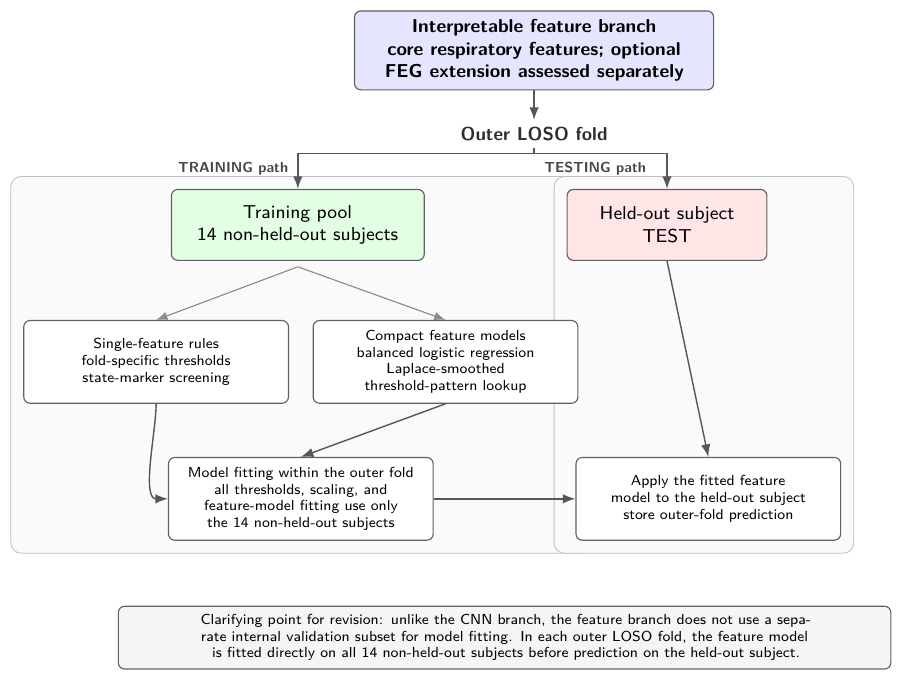}
\caption{{Outer-LOSO training and testing path for the interpretable feature branch. The held-out subject is used only for the outer TEST evaluation. For the fitted feature model, fold-specific thresholds, scaling/imputation quantities, Laplace lookup tables or continuous classifier parameters are obtained from the 14 non-held-out subjects. The exploratory candidate-combination search is described separately in the text and is not represented as nested selection in this diagram.}}
\label{fig:feature_loso_workflow}
\end{figure}

{Figure~\ref{fig:feature_loso_workflow} makes the outer-fold fitting/evaluation path of the feature branch explicit. For a given outer fold and a given candidate feature model, all fold-specific thresholds, imputation/scaling statistics, lookup-table quantities, score thresholds, and classifier parameters are estimated from the 14 non-held-out subjects, after which the fitted model is applied to the held-out subject. This diagram concerns fold-wise model fitting and testing; the exploratory, non-nested candidate-combination search described above is a separate selection layer.\par}

\FloatBarrier

\subsection{Compact 1D-CNN model}\label{subsec:cnn}

The raw-signal neural model was a compact one-dimensional convolutional neural network (1D-CNN-Tiny). The model receives a normalized 60 s RESP segment of length 3000 as input and uses four convolutional stages with channel configuration [16, 32, 64, 128], followed by a lightweight classification head. The total number of trainable parameters is approximately 0.229M. The model was optimized using AdamW with learning rate $3\times10^{-4}$, weight decay $10^{-4}$, dropout 0.20, batch size 64, and random seed 42. {Within each outer LOSO fold, the 14 non-held-out subjects formed the CNN training pool. An internal TRAIN/VAL split within this pool was used only for training-time selection (best epoch). After this selection, a newly initialized CNN was trained again on the complete 14-subject outer-training pool using the selected training duration and was then evaluated on the held-out subject.\par}

{Unlike the CNN branch, the compact feature models did not require a separate final TRAIN+VAL refit because they did not use a separate internal validation subset for model fitting. For every outer LOSO fold, fold-specific feature thresholds, imputation and scaling statistics, Laplace lookup tables, score thresholds, and continuous-feature classifier parameters were fitted directly using the complete set of 14 non-held-out subjects and were then applied unchanged to the held-out subject. Thus, the final CNN and feature-model predictions in each outer fold were both produced from models fitted using all 14 non-held-out subjects; the branches differ in their internal optimization procedure, not in the amount of data used for the final outer-fold model fit. The exploratory selection of feature combinations is a separate issue and is described as non-nested in Section~\ref{subsec:feature_models}.\par}

\begin{figure}[!htbp]
\centering
\includegraphics[width=0.80\textwidth]{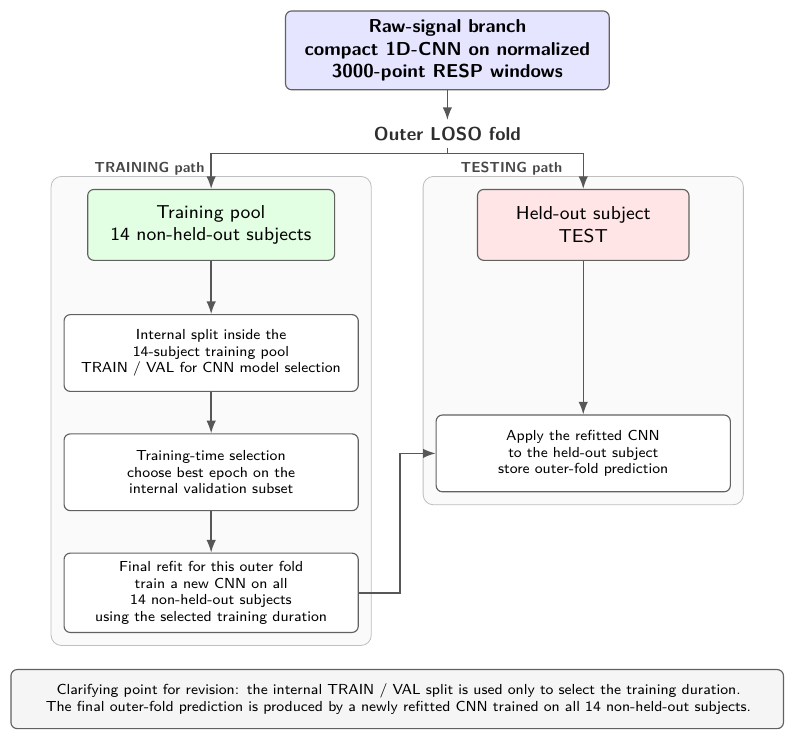}
\caption{{Outer-LOSO training and testing path for the raw-signal CNN branch. The 14 non-held-out subjects form the outer training pool. An internal TRAIN/VAL split is used only for training-time selection; a newly initialized CNN is then refitted on the complete 14-subject pool using the selected training duration and finally applied to the held-out subject.}}
\label{fig:cnn_loso_workflow}
\end{figure}

{The corresponding CNN path is shown in Fig.~\ref{fig:cnn_loso_workflow}. The diagram separates the internal TRAIN/VAL step used to determine training duration from the final outer-fold refit and the held-out TEST evaluation.\par}

\FloatBarrier

\subsection{Evaluation metrics and comparison strategy}\label{subsec:evaluation}

Performance was evaluated using accuracy, macro-F1, positive-class recall, positive-class precision, balanced accuracy, and Matthews correlation coefficient (MCC). MCC was used as the primary ranking criterion because it accounts for all four confusion-matrix entries and is suitable for imbalanced binary tasks. The one-vs-rest analysis used the same metrics, with the selected state treated as the positive class.

The results are interpreted at three levels. First, single-feature thresholds identify individual respiratory markers. Second, compact binary-pattern and balanced-logistic-regression models identify small state-specific feature combinations. Third, the CNN provides a raw-window predictive model. The comparison between core interpretable features and the full FEG-Pro feature set is reported separately to determine whether complex nonlinear forecast-error-growth descriptors materially change the conclusions.

\clearpage

\section{Results}\label{sec:results}

\subsection{Single-feature one-vs-rest threshold rules}\label{subsec:single_threshold_results}

The first result layer evaluates whether individual respiratory descriptors can separate each WESAD state from all remaining states. For each feature and each held-out subject, the threshold and its direction were selected using only the training subjects. Table~\ref{tab:single_ovr_rules} reports the strongest single-feature rule for each one-vs-rest state. The threshold value is reported as the mean across the 15 LOSO folds with the corresponding standard deviation. The column ``positive side'' indicates whether the target state was predicted on the left side of the threshold, $f\leq\bar{\theta}$, or on the right side, $f\geq\bar{\theta}$. For all four best rules, the same direction was selected in all 15 folds.

\begin{table}[htbp]
\caption{Best single-feature one-vs-rest threshold rules. Thresholds are means across 15 LOSO folds. Metrics are shown as percentages.}\label{tab:single_ovr_rules}%
\centering
\scriptsize
\begin{tabularx}{\textwidth}{@{}p{0.12\textwidth}p{0.22\textwidth}p{0.12\textwidth}Yrrr@{}}
\toprule
Target state & Best marker & Group / level & Positive rule & MCC & Macro-F1 & Recall \\
\midrule
Baseline & Zmp (WW) & ANP & Left: $f\leq180.60\pm0.91$ & 57.97 & 77.17 & 82.55 \\
Stress & Fre6 (0.70--1.00 Hz) & STF & Right: $f\geq139.32\pm4.15$ & 72.70 & 84.94 & 85.01 \\
Amusement & BR & RT & Right: $f\geq17.75\pm0.06$ & 28.19 & 59.62 & 54.97 \\
Meditation & Zmp (BC-median) & ANP & Right: $f\geq241.49\pm0.64$ & 79.82 & 88.46 & 78.62 \\
\botrule
\end{tabularx}
\footnotetext{RT: respiratory timing; STF: spectral/time-frequency; ANP: autocorrelation and nonlinear predictability; WW: whole window; BC: breath-cycle aggregation. Zmp is the autocorrelation negative-to-positive transition lag. Thresholds are reported in native feature units.}
\end{table}

The one-vs-rest threshold screening already reveals state-specific structure. Baseline was best identified by a shorter whole-window Zmp autocorrelation transition lag. Stress was best identified by high-frequency respiratory spectral energy (Fre6), with the target class on the high-value side of the threshold. Amusement was weakly separable from RESP alone; its best single marker was breathing rate, but the MCC remained low. Meditation showed the strongest single-feature separability after stress, with high values of the breath-cycle median Zmp lag marking the meditation class. These findings support the later interpretation that stress and meditation have more stable respiratory signatures than amusement in the RESP-only setting.

\subsection{Single-feature stress markers}\label{subsec:single_features}

\begin{figure}[htbp]
\centering
\begin{minipage}{0.49\textwidth}
\includegraphics[width=\linewidth]{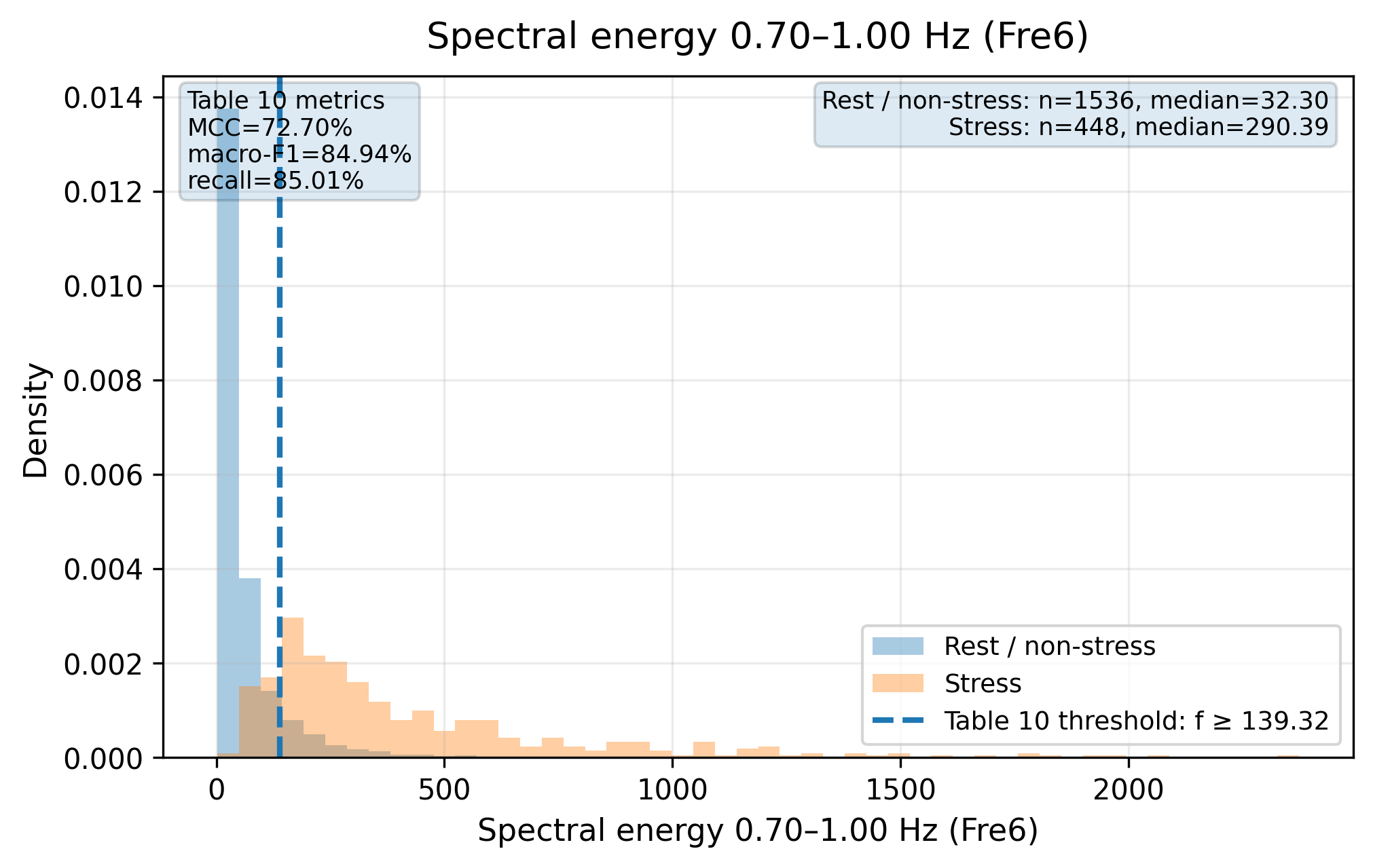}\\[-2pt]
\small (a) Spectral energy 0.70--1.00 Hz (Fre6).
\end{minipage}\hfill
\begin{minipage}{0.49\textwidth}
\includegraphics[width=\linewidth]{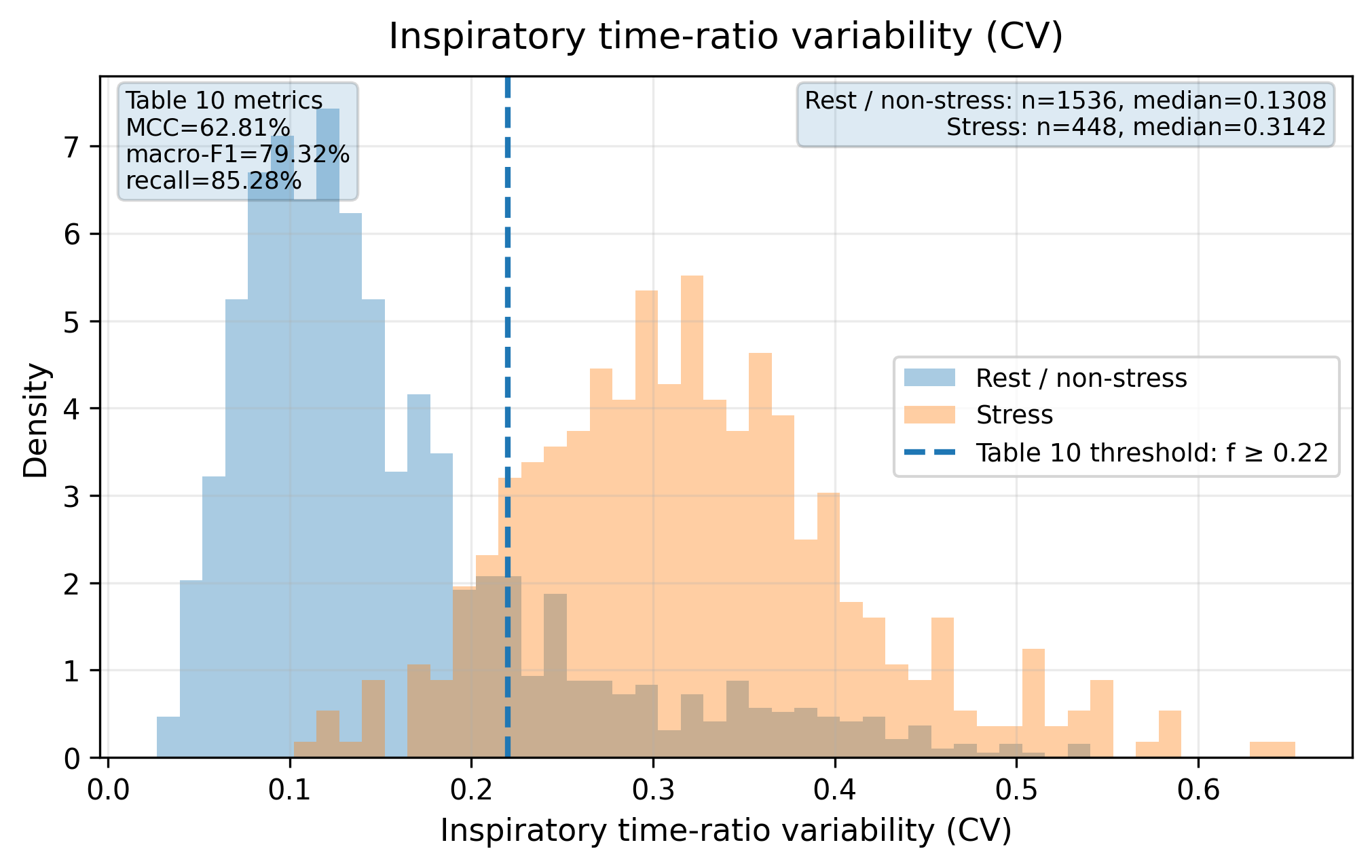}\\[-2pt]
\small (b) Inspiratory time-ratio variability.
\end{minipage}

\begin{minipage}{0.49\textwidth}
\includegraphics[width=\linewidth]{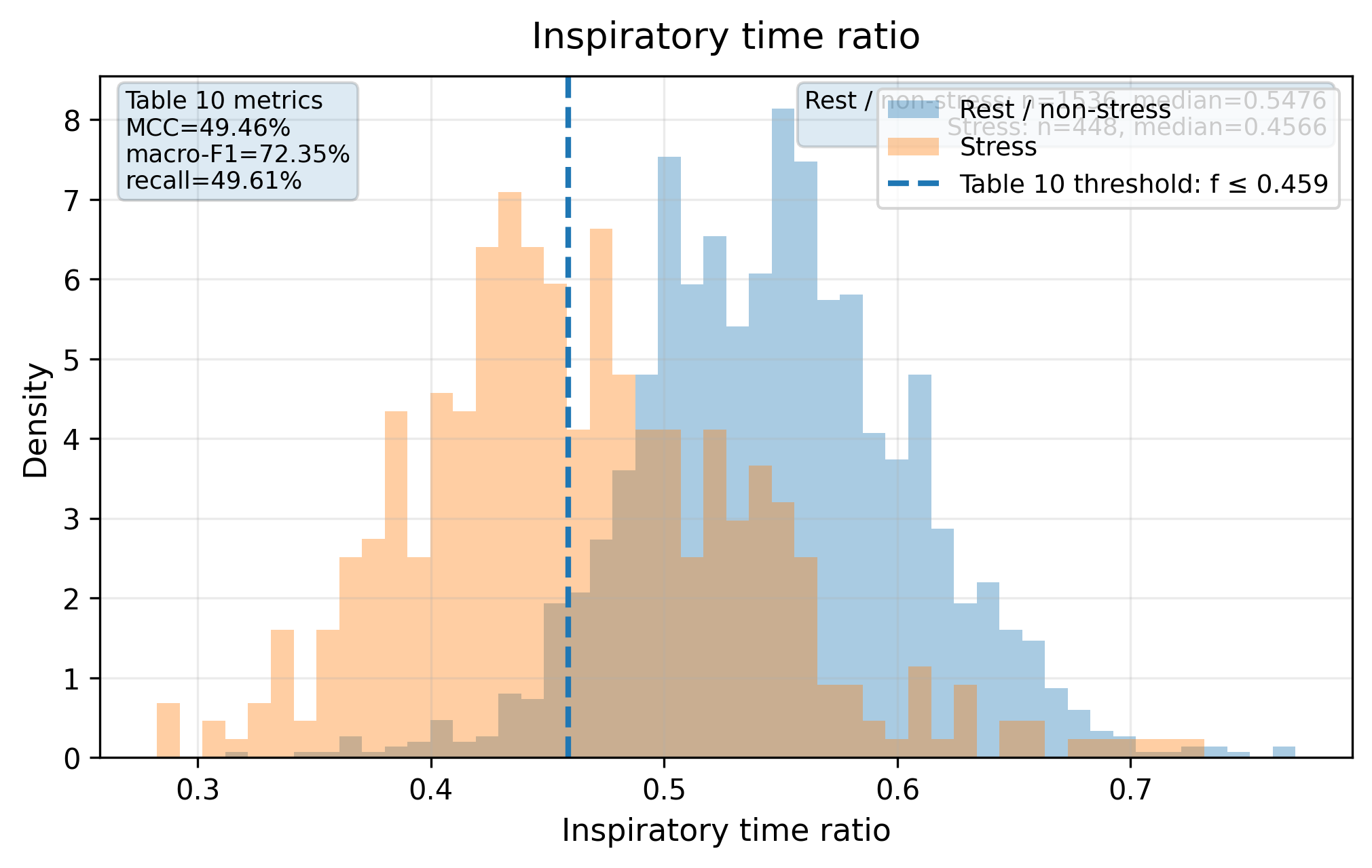}\\[-2pt]
\small (c) Inspiratory time ratio.
\end{minipage}\hfill
\begin{minipage}{0.49\textwidth}
\includegraphics[width=\linewidth]{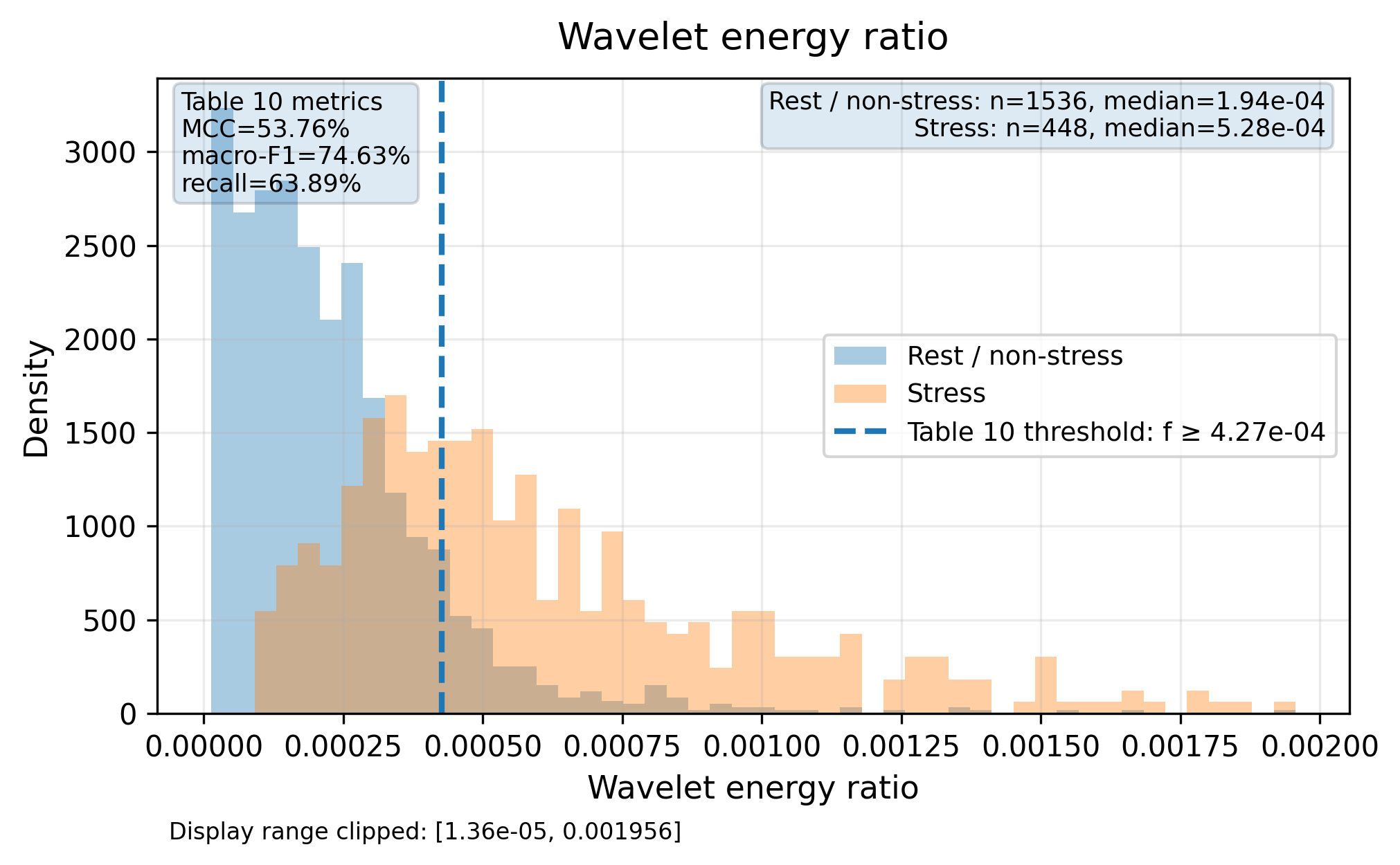}\\[-2pt]
\small (d) Wavelet energy ratio.
\end{minipage}
\caption{Representative distributions of selected single-feature stress markers used in Table~\ref{tab:stress_single_thresholds}. The histograms compare stress windows with all non-stress windows; the dashed vertical lines indicate the mean LOSO thresholds from Table~\ref{tab:stress_single_thresholds}.}
\label{fig:feature_distributions}
\end{figure}

Because stress detection is the primary practical task, Table~\ref{tab:stress_single_thresholds} lists the leading and representative single-feature threshold rules for stress-vs-rest detection. The table includes the best overall stress marker, selected exploratory nonlinear descriptors, and the core physiological/signal-processing markers shown in Fig.~\ref{fig:feature_distributions}. The best single marker was Fre6, the respiratory spectral energy in the 0.70--1.00 Hz band, with MCC=72.70. Among the core physiological descriptors, inspiratory time-ratio variability (IT\_ratio\_cv) remained the strongest breath-cycle timing marker. The leading nonlinear rules were not direct clinical respiratory measures; they summarized breath-cycle variability of forecast-error-growth profiles and are therefore interpreted as exploratory predictability markers.

\begin{table}[htbp]
\caption{Representative single-feature threshold rules for stress-vs-rest detection. Thresholds are means across 15 LOSO folds. Metrics are shown as percentages.}\label{tab:stress_single_thresholds}%
\centering
\scriptsize
\begin{tabularx}{\textwidth}{@{}p{0.20\textwidth}p{0.10\textwidth}p{0.09\textwidth}Yrrr@{}}
\toprule
Stress marker & Group & Type & Positive rule & MCC & Macro-F1 & Recall \\
\midrule
Fre6 & STF & Core & Right: $f\geq139.32\pm4.15$ & 72.70 & 84.94 & 85.01 \\
BC-CV slope\_two\_left & ANP-FEG & FEG-Pro & Right: $f\geq0.945\pm0.087$ & 66.08 & 82.24 & 75.59 \\
BC-median FEDE delta & ANP-FEG & FEG-Pro & Left: $f\leq0.01599\pm0.00045$ & 64.69 & 81.76 & 71.47 \\
BC-mean roughness & ANP-FEG & FEG-Pro & Right: $f\geq0.06355\pm0.00066$ & 63.07 & 80.53 & 78.83 \\
IT\_ratio\_cv & BV & Core & Right: $f\geq0.220\pm0.012$ & 62.81 & 79.32 & 85.28 \\
Fre5 & STF & Core & Right: $f\geq1236.11\pm35.97$ & 61.64 & 78.53 & 80.96 \\
Wer & STF & Core & Right: $f\geq4.27\times10^{-4}\pm1.5\times10^{-6}$ & 53.76 & 74.63 & 63.89 \\
IT\_ratio & RT & Core & Left: $f\leq0.459\pm0.010$ & 49.46 & 72.35 & 49.61 \\
Skewness & WS & Core & Right: $f\geq0.594\pm0.013$ & 49.75 & 73.20 & 48.63 \\
\botrule
\end{tabularx}
\footnotetext{BV: breath-to-breath variability; RT: respiratory timing; STF: spectral/time-frequency; WS: waveform statistics; ANP-FEG: exploratory forecast-error-growth descriptor within the ANP group. The Type column separates the core interpretable feature set from exploratory FEG-Pro descriptors.}
\end{table}

The distribution plots in Fig.~\ref{fig:feature_distributions} illustrate how the strongest spectral and physiological markers separate rest/non-stress and stress. For visualization, long distribution tails were clipped only in the displayed range; all metrics were computed on the full data. The plots are used only as a visual explanation of the threshold rules; the reported thresholds and metrics were obtained from fold-wise training-subject optimization under LOSO validation.

\FloatBarrier

\subsection{Compact core feature combinations}\label{subsec:compact_core_combinations}

The second result layer evaluates whether small combinations of respiratory descriptors improve over the single-feature rules. In this subsection, only the core interpretable feature set is used: respiratory timing, breath-to-breath variability, waveform statistics, spectral/time-frequency descriptors, and autocorrelation transition-lag descriptors. The exploratory forecast-error-growth profile descriptors are excluded here. This separation is important because the aim of the core analysis is to identify respiratory signatures that remain physiologically transparent.

\begin{figure}[htbp]
\centering
\includegraphics[width=0.98\textwidth]{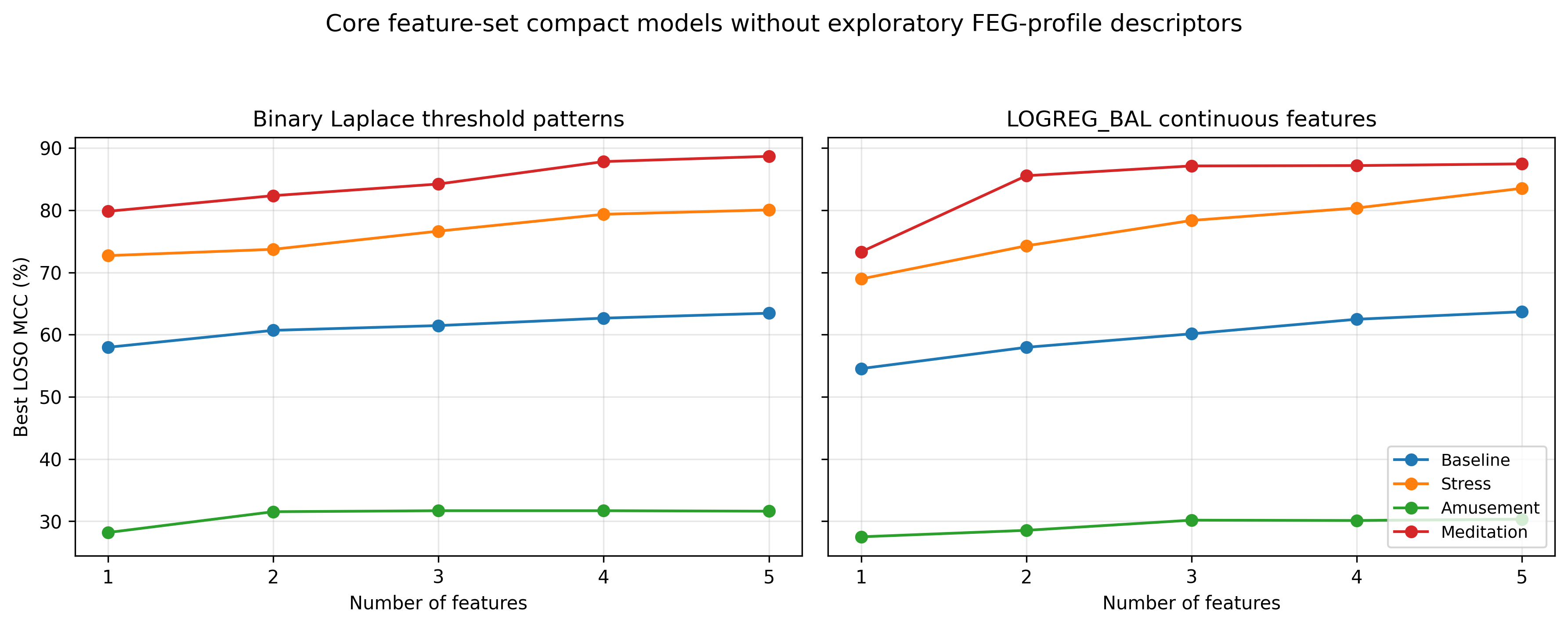}
\caption{Best LOSO MCC as a function of the number of input features for compact models based on the core feature set only. Exploratory forecast-error-growth profile descriptors are excluded. The binary Laplace model uses fold-specific threshold indicators, whereas balanced logistic regression uses continuous feature values.}
\label{fig:combo_core_mcc}
\end{figure}

Two compact model families were compared. The binary Laplace threshold-pattern model first converted each selected feature into a fold-specific binary rule and then modeled the resulting binary pattern. The balanced-logistic-regression model used the continuous feature values directly. Figure~\ref{fig:combo_core_mcc} summarizes how the best LOSO MCC changed with the number of selected features for both models and all four one-vs-rest states.

Table~\ref{tab:core_combo_best} lists the best core-feature combination found for each target state and model family. The results show a clear state-dependent pattern. For stress, balanced logistic regression provided the strongest core-feature model, reaching MCC=83.51 with five continuous features. The selected stress signature combined breath-cycle variability of the Zmp autocorrelation transition lag, wavelet energy ratio, waveform skewness, inspiratory time ratio, and signal energy. In contrast, meditation was best captured by the binary Laplace threshold-pattern model, reaching MCC=88.65 using a combination of breath-cycle median Zmp, low-frequency FFT power, wavelet energy entropy, Fre5, and inspiratory time ratio. Baseline showed moderate separability for both models, whereas amusement remained weakly separable from RESP alone.

\begin{table}[htbp]
\caption{Best compact feature combinations using the core interpretable feature set only. Exploratory forecast-error-growth profile descriptors are excluded. Metrics are shown as percentages.}\label{tab:core_combo_best}%
\centering
\scriptsize
\begin{tabularx}{\textwidth}{@{}p{0.105\textwidth}c c r r Y@{}}
\toprule
Target & Model & $k$ & MCC & Macro-F1 & Selected feature combination \\
\midrule
Baseline & BL & 5 & 63.44 & 81.72 & Fre (dominant); SD1; IT ratio CV; SampEn; signal SD \\
Baseline & LR & 5 & 63.67 & 81.16 & FFT 0.20-0.50 Hz; SEG3-mean Zmp; SEG3-SD Zmp; SampEn; IT CV \\
Stress & BL & 5 & 80.04 & 90.00 & IT ratio CV; Wer; skewness; BC-IQR Zpm; FFT 0.50-1.00 Hz \\
Stress & LR & 5 & 83.51 & 91.46 & BC-CV Zmp; Wer; skewness; IT ratio; energy \\
Amusement & BL & 3 & 31.70 & 60.64 & BC-mean Zpm; Fre (dominant); Fre6 \\
Amusement & LR & 5 & 30.34 & 55.20 & BC-median Zpm; WW Zmp; ET; Fre (dominant); ZCR \\
Meditation & BL & 5 & 88.65 & 94.20 & BC-median Zmp; FFT 0.05-0.20 Hz; Wee; Fre5; IT ratio \\
Meditation & LR & 5 & 87.44 & 93.71 & BC-mean Zmp; Wee; BC-median Zpm; Fre5; Fre2 \\
\botrule
\end{tabularx}
\footnotetext{BL: binary Laplace threshold-pattern model; LR: balanced logistic regression. Fre: dominant respiratory frequency; Fre2/Fre5/Fre6: Yang-style frequency-band energies; FFT values denote whole-window FFT band powers; BC denotes breath-cycle aggregation; SEG3 denotes three-segment aggregation; Zpm/Zmp are autocorrelation transition-lag descriptors.}
\end{table}

The model comparison in Table~\ref{tab:core_combo_best} also indicates why both compact approaches are useful. Binary threshold patterns are especially competitive for meditation and for low-dimensional baseline rules, because discretized respiratory states can capture nonlinear rule-like behavior. Balanced logistic regression is stronger for stress, where continuous variation in waveform asymmetry, energy, inspiratory timing, and autocorrelation-lag variability appears to be informative. Thus, the two models are not redundant: threshold rules provide simple interpretable decision patterns, whereas balanced logistic regression provides a stronger continuous-feature classifier for the primary stress-detection problem.

{
\subsection{Nested sensitivity analysis of the Top-20 stress search}\label{subsec:nested_top20_sensitivity}

Because the auxiliary Top-20 search described in Section~\ref{subsec:feature_models} used non-nested global ranking, we repeated the stress analysis with the strict nested procedure described above. Table~\ref{tab:nested_top20_sensitivity} compares the original exploratory LOSO estimate with the fully isolated outer-LOSO result. The mean MCC decreased from 85.45\% to 82.29\%, a difference of 3.16 percentage points. Macro-F1, stress recall, and accuracy decreased by 2.23, 2.01, and 2.00 percentage points, respectively. Thus, the original screening introduced a modest optimistic bias, but the conclusion that compact respiratory features carry strong subject-independent stress information was preserved.

\begin{table}[htbp]
\caption{Sensitivity of the auxiliary 52-feature Top-20 stress search to strict nested feature selection. All metrics are percentages and are means across subject-wise outer folds.}\label{tab:nested_top20_sensitivity}%
\centering
\scriptsize
\begin{tabularx}{\textwidth}{@{}Yrrrr@{}}
\toprule
Validation procedure & MCC & Macro-F1 & Stress recall & Accuracy \\
\midrule
Original exploratory Top-20 LOSO & 85.45 & 92.04 & 92.86 & 93.87 \\
Strict nested Top-20 LOSO & 82.29 & 89.81 & 90.85 & 91.88 \\
Difference (nested $-$ exploratory) & $-3.16$ & $-2.23$ & $-2.01$ & $-2.00$ \\
\bottomrule
\end{tabularx}
\footnotetext{The exploratory row used the original global Top-20 ranking. In the nested row, each outer test subject was excluded before feature ranking and subset selection; inner validation used five subject-wise GroupKFold splits, exhaustive four-feature evaluation, and deterministic beam expansion for sizes 5--7.}
\end{table}

The nested selections were also structurally stable across outer folds. Inspiratory time ratio was selected in all 15 folds, waveform standard deviation and skewness in 14/15 folds, and wavelet energy ratio in 11/15 folds. These recurrence patterns support the robustness of inspiratory timing, waveform-shape/dispersion, and time-frequency energy families even when feature ranking and subset selection are fully isolated from the held-out subject.
\par}

\subsection{Effect of exploratory FEG-profile descriptors}\label{subsec:feg_extension_results}

After establishing the core signatures, we evaluated the extended feature set that additionally includes exploratory forecast-error-growth (FEG-profile) descriptors. This extension tests whether more complex nonlinear predictability summaries materially improve the compact models. Table~\ref{tab:feg_extension_best} reports, for each target state, the best result obtained after adding the full FEG-profile descriptor set and compares it with the best core-feature result for the same state.

\begin{table}[htbp]
\caption{Best results after adding the full exploratory FEG-profile descriptor set. The core reference is the best result from Table~\ref{tab:core_combo_best} for the same target state. Metrics are shown as MCC percentages.}\label{tab:feg_extension_best}%
\centering
\scriptsize
\begin{tabularx}{\textwidth}{@{}p{0.105\textwidth}c c r r r Y@{}}
\toprule
Target & Model & $k$ & Full-FEG MCC & Core MCC & $\Delta$MCC & Selected full-FEG combination \\
\midrule
Baseline & LR & 5 & 65.34 & 63.67 & +1.67 & FFT 0.20-0.50 Hz; SEG3-mean Zmp; SEG3-SD Zmp; BC-median FEDE early slope; SampEn \\
Stress & LR & 5 & 81.02 & 83.51 & -2.49 & Fre6; BC-median FEDE delta; BC-IQR two-line R2; BC-IQR residual roughness; BC-IQR global FEG slope \\
Amusement & BL & 5 & 35.69 & 31.70 & +3.99 & BC-median Zpm-used; BC-mean Zpm; FFT 0.50-1.00 Hz; Fre6; BC-median residual curvature \\
Meditation & BL & 5 & 87.36 & 88.65 & -1.29 & BC-median Zmp; Wer; FFT 0.05-0.20 Hz; BC-range right FEG slope; BC-range FEDE mean \\
\botrule
\end{tabularx}
\footnotetext{BL: binary Laplace threshold-pattern model; LR: balanced logistic regression. Positive $\Delta$MCC indicates improvement relative to the best core-feature model for the same target state. FEG-profile features include slope, FEDE, residual-shape, roughness, monotonicity, and local $R^2$ descriptors summarized in Table~\ref{tab:feg_descriptors}.}
\end{table}

The full FEG-profile extension produced a modest gain for baseline (+1.67 MCC points) and a larger but still low-absolute-performance gain for amusement (+3.99 MCC points). However, it did not improve the two most physiologically informative states. For stress, the best full-FEG model reached MCC=81.02, below the core balanced-logistic-regression result of 83.51. For meditation, the best full-FEG result reached MCC=87.36, below the core binary Laplace result of 88.65. Therefore, the main physiological conclusions are based on the core interpretable descriptors, whereas the FEG-profile descriptors are treated as a nonlinear exploratory extension that may help in selected cases but is not necessary for the main stress and meditation signatures.

\subsection{Raw-signal CNN final refit and comparison with feature models}\label{subsec:cnn_final_comparison}

The compact 1D-CNN was evaluated on the same four one-vs-rest tasks using the final train+validation refit protocol. To keep the comparison unambiguous, only the final refit results are reported in this subsection. Table~\ref{tab:cnn_refit_results} summarizes the pooled LOSO metrics obtained from the summed confusion matrices across the 15 held-out subjects. The CNN was strongest for the practical stress-vs-rest detector, reaching accuracy 96.72\%, macro-F1 95.30\%, and MCC 90.61\%. The same model also produced high absolute accuracy for meditation, but its MCC and macro-F1 were below the best compact feature model for that state.

\begin{table}[htbp]
\caption{Final train+validation refit performance of the compact raw-signal 1D-CNN. Metrics are pooled LOSO values from the summed confusion matrix and are shown as percentages. Target precision, recall, and F1 refer to the positive one-vs-rest state.}\label{tab:cnn_refit_results}%
\centering
\scriptsize
\setlength{\tabcolsep}{2.2pt}%
\begin{tabular}{@{}lrrrrrrrr@{}}
\toprule
Target & Acc. & Macro P & Macro R & Macro-F1 & MCC & Target P & Target R & Target F1 \\
\midrule
Baseline & 77.07 & 76.57 & 75.99 & 76.22 & 52.56 & 74.26 & 69.35 & 71.72 \\
Stress & 96.72 & 95.41 & 95.20 & 95.30 & 90.61 & 93.03 & 92.41 & 92.72 \\
Amusement & 86.14 & 64.78 & 63.45 & 64.06 & 28.20 & 37.93 & 34.07 & 35.90 \\
Meditation & 91.38 & 87.38 & 90.25 & 88.67 & 77.58 & 78.69 & 88.08 & 83.12 \\
\botrule
\end{tabular}
\end{table}

For a direct comparison with the interpretable branch, Table~\ref{tab:best_feature_all_metrics} reports the best compact feature model found for each target state across the core and full-FEG feature sets, while Table~\ref{tab:best_feature_signatures} lists the corresponding feature signatures. The comparison deliberately uses the best available compact feature model for each state rather than forcing a single feature-model family across all states, because the earlier analysis showed that binary threshold patterns and balanced logistic regression capture different forms of respiratory structure.

\begin{table}[htbp]
\caption{Best compact feature-model performance for each one-vs-rest state. The best model was selected by MCC from the core-feature and full-FEG searches. Metrics are pooled LOSO values and are shown as percentages. Target precision, recall, and F1 refer to the positive one-vs-rest state.}\label{tab:best_feature_all_metrics}%
\centering
\scriptsize
\setlength{\tabcolsep}{1.6pt}%
\begin{tabular}{@{}lllcrrrrrr@{}}
\toprule
Target & Model & Set & $k$ & Acc. & Macro-F1 & MCC & Target P & Target R & Target F1 \\
\midrule
Baseline & LR & Full-FEG & 5 & 82.26 & 82.14 & 65.34 & 74.24 & 88.34 & 80.68 \\
Stress & LR & Core & 5 & 93.65 & 91.46 & 83.51 & 80.26 & 95.31 & 87.14 \\
Amusement & BL & Full-FEG & 5 & 78.38 & 64.55 & 35.69 & 30.44 & 69.91 & 42.42 \\
Meditation & BL & Core & 5 & 95.92 & 94.20 & 88.65 & 96.49 & 86.19 & 91.05 \\
\botrule
\end{tabular}
\footnotetext{BL: binary Laplace threshold-pattern model; LR: balanced logistic regression. Core: interpretable feature set excluding exploratory FEG-profile descriptors; Full-FEG: extended feature set including exploratory forecast-error-growth descriptors.}
\end{table}

\begin{table}[htbp]
\caption{Selected feature signatures of the best compact feature models in Table~\ref{tab:best_feature_all_metrics}.}\label{tab:best_feature_signatures}%
\centering
\scriptsize
\begin{tabularx}{\textwidth}{@{}p{0.14\textwidth}p{0.12\textwidth}Y@{}}
\toprule
Target & Model & Feature signature \\
\midrule
Baseline & LR & FFT 0.20-0.50 Hz; SEG3-mean Zmp; SEG3-SD Zmp; BC-median FEDE early slope; SampEn \\
Stress & LR & BC-CV Zmp; Wer; skewness; IT ratio; energy \\
Amusement & BL & BC-median Zpm-used; BC-mean Zpm; FFT 0.50-1.00 Hz; Fre6; BC-median residual curvature \\
Meditation & BL & BC-median Zmp; FFT 0.05-0.20 Hz; Wee; Fre5; IT ratio \\
\botrule
\end{tabularx}
\end{table}

\begin{figure}[htbp]
\centering
\includegraphics[width=0.85\textwidth]{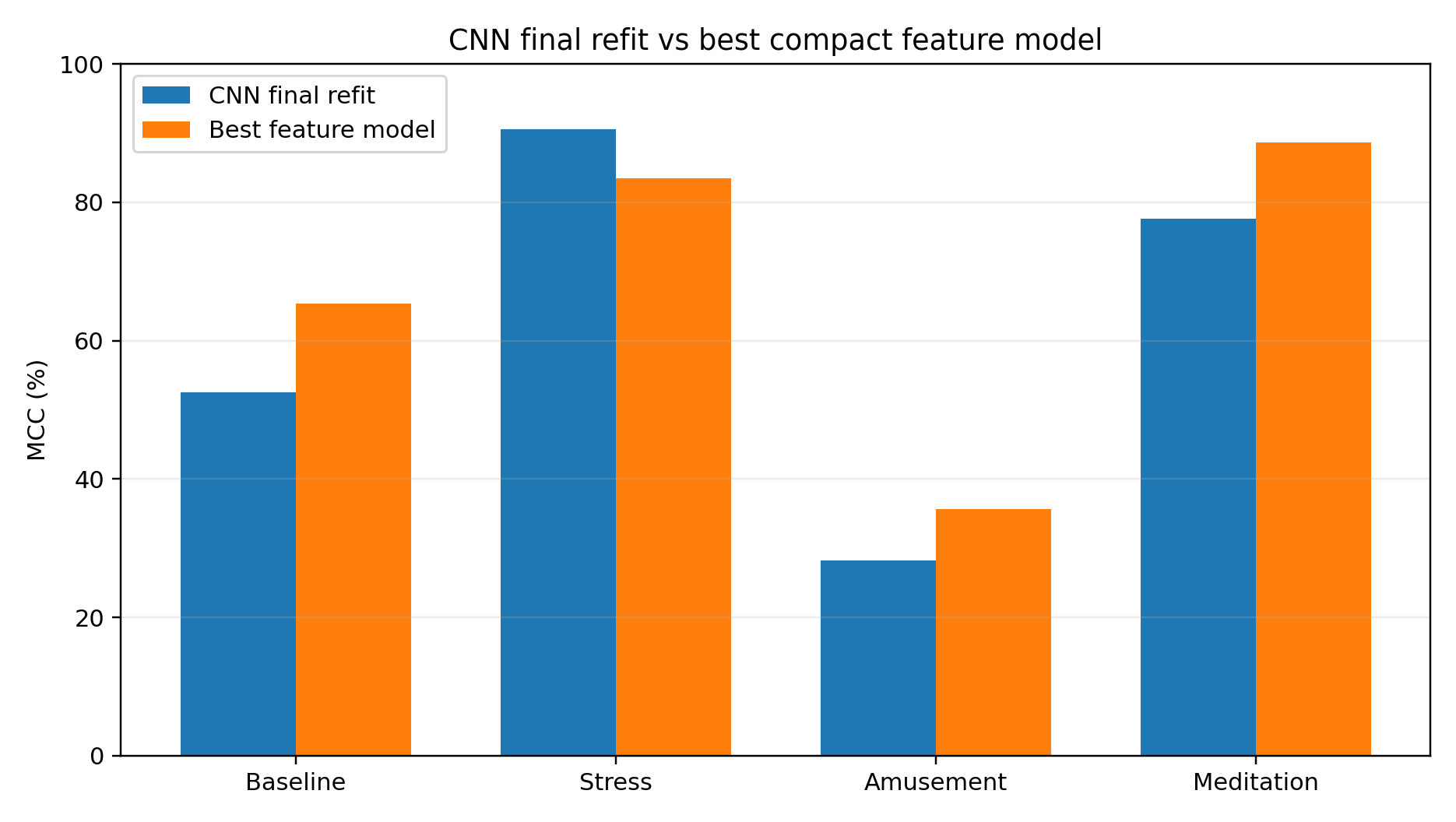}
\caption{Final comparison between the raw-signal CNN refit and the best compact feature model for each one-vs-rest state. Bars show pooled LOSO MCC.}
\label{fig:cnn_feature_comparison}
\end{figure}

Figure~\ref{fig:cnn_feature_comparison} and Table~\ref{tab:cnn_feature_comparison} summarize the final branch-level comparison. The CNN final refit clearly outperformed the best compact feature model for stress, improving MCC by 7.10 points. However, the best compact feature model remained stronger for baseline, amusement, and meditation. The largest advantage of the feature branch was observed for meditation, where the binary threshold-pattern model achieved MCC 88.65 compared with 77.58 for the final-refit CNN. Thus, the CNN branch is most useful as a high-performance stress detector, whereas compact feature signatures remain preferable for interpreting and in some cases predicting state-specific respiratory structure.

\begin{table}[htbp]
\caption{Final MCC comparison between the raw-signal CNN and the best compact feature model. Values are pooled LOSO MCC percentages.}\label{tab:cnn_feature_comparison}%
\centering
\scriptsize
\begin{tabularx}{\textwidth}{@{}p{0.14\textwidth}rrrY@{}}
\toprule
Target & CNN & Best feature & CNN - feature & Stronger branch \\
\midrule
Baseline & 52.56 & 65.34 & -12.78 & Feature model \\
Stress & 90.61 & 83.51 & +7.10 & CNN \\
Amusement & 28.20 & 35.69 & -7.49 & Feature model \\
Meditation & 77.58 & 88.65 & -11.07 & Feature model \\
\botrule
\end{tabularx}
\end{table}

\FloatBarrier

\section{Discussion}\label{sec:discussion}

The results support the main rationale of the study: respiratory affect recognition should be analyzed not only as a classification problem but also as a state-specific signature problem. The strongest single stress marker was Fre6, indicating that stress increased respiratory spectral energy in the 0.70--1.00 Hz band. However, the best compact stress model required several complementary components: autocorrelation-lag variability, wavelet energy, waveform skewness, inspiratory timing, and signal energy. This suggests that stress is not represented by a single respiratory property, but by a combined change in rhythm, waveform asymmetry, phase organization, and energy distribution.

This pattern is consistent with the broader psychophysiological view that stress modifies both respiratory rhythm and cardiorespiratory spectral structure. Respiratory-frequency information has been shown to improve stress assessment when interpreting heart-rate variability, because breathing rate can shift power between conventional frequency bands and alter the apparent autonomic balance \cite{hernando2016resp_freq}. Recent controlled comparisons between paced breathing and stressful mental states also support the idea that stress is reflected by a joint change in respiratory behavior and linear/nonlinear variability rather than by a single isolated scalar \cite{chand2024paced_stress}. The result is also compatible with evidence that anxiety and stress can alter respiratory rate variability, inspiratory/expiratory timing, and breathing complexity \cite{ritsert2022anxiety,suzuki2023resp_params,tiwari2019complexity}. Therefore, the Fre6 marker should be interpreted as a high-frequency respiratory-energy signature of stress-like breathing within the present windowing and sampling scheme, not as a universal clinical threshold.

The final CNN comparison adds an important predictive layer to this interpretation. The compact raw-signal CNN was clearly strongest for the practical stress-vs-rest detector after final refit, reaching MCC=90.61, macro-F1=95.30\%, and accuracy=96.72\%. This result indicates that the raw respiratory waveform contains stress-related temporal information that is only partially captured by the handcrafted feature set. Nevertheless, the best compact stress feature model still reached MCC=83.51 and had a very high stress recall of 95.31\%, making it useful as an interpretable stress signature even when it is not the strongest predictive model. This finding fits the broader pattern in which multimodal and deep models often maximize accuracy, whereas interpretable or lightweight models provide a clearer path to biomarker attribution, resource-limited deployment, and clinical explanation \cite{bolpagni2024personalized,chatzaki2025overview,ghose2025lightweight}.

{The additional nested Top-20 sensitivity analysis strengthens this interpretation by separating predictive generalization from exploratory subset discovery. Strict nesting reduced the auxiliary stress-search MCC from 85.45\% to 82.29\%, indicating modest selection optimism rather than a collapse of performance; the repeated selection of inspiratory timing, waveform dispersion/skewness, and wavelet-energy variables across outer folds further supports the stability of these feature families.\par}

The non-stress state signatures behaved differently. For meditation, the best compact feature model outperformed the final-refit CNN by a large margin (MCC=88.65 vs 77.58). This supports the interpretation that meditation is well described by explicit respiratory structure: breath-cycle median Zmp, low-frequency FFT power, wavelet energy entropy, Fre5, and inspiratory timing. The feature branch also outperformed the CNN for baseline and amusement in terms of MCC. Therefore, the CNN should not be interpreted as universally superior for all RESP-based state separation tasks. Instead, it appears most beneficial for the stress detector, whereas handcrafted signatures remain more effective for several state-specific comparisons.

The meditation result is also physiologically plausible. Slow-breathing and meditation studies report lower respiratory frequency, altered phase timing, and changes in breath-rate variability or regularity \cite{zaccaro2018slowbreathing,kral2023slower,soni2019breath_variability}. Breathwork trials and reviews further indicate that structured breathing can reduce physiological arousal and improve mood or stress-related outcomes, although the exact respiratory pattern depends on the technique \cite{balban2023breathwork,fincham2023breathwork,birdee2023slowbreathing}. Experimental work on inhalation/exhalation ratio indicates that phase balance can influence subjective relaxation and respiratory sinus arrhythmia, which supports the relevance of inspiratory timing and phase-balance descriptors \cite{diest2014inhalation}. In this context, the combination of low-frequency FFT power, inspiratory timing, wavelet entropy, and Zmp transition scale can be interpreted as a compact signature of slower, more structured, and more temporally persistent breathing during meditation.

{
Amusement remained the least separable state from RESP alone. This result is consistent with the relatively small respiratory contrast between amusement and resting conditions. In our data, baseline and amusement showed almost identical mean respiratory cycle rates (15.09 $\pm$ 3.08 and 15.54 $\pm$ 3.73 cycles/min, respectively), indicating substantial overlap already at the level of basic respiratory timing. Previous psychophysiological studies similarly suggest that amusement produces comparatively subtle respiratory changes, including reductions in inspiratory time and tidal volume, rather than the pronounced respiratory reorganization observed for more strongly arousing emotional conditions \cite{boiten1998emotional}. More generally, respiratory variables have been reported to track affective arousal more consistently than stimulus pleasantness or valence; high-arousal stimuli produce clearer changes in ventilation, inspiratory drive, and respiratory timing, whereas pleasantness effects are weaker or less consistent \cite{gomez2005respiratory,gomez2008respiration}. This interpretation is also compatible with respiration-only emotion-recognition studies in which arousal discrimination was stronger than valence discrimination \cite{zhang2017respiration}. Finally, inter-subject variability can further reduce generalized affect-recognition performance relative to personalized models \cite{li2023personalized}. Taken together, the low amusement MCC in the present LOSO analysis is therefore likely to reflect a genuinely low-contrast and subject-dependent respiratory signature rather than simply insufficient classifier complexity. Multimodal physiological information may be required when reliable discrimination of amusement from resting or other low-contrast affective states is the primary objective.
\par}

{
More broadly, recent confusion-aware recognition approaches emphasize that strong aggregate performance can conceal substantial class-specific failure modes \cite{zhu2026confusion}. This consideration is particularly relevant here because the high performance of the binary stress detector does not imply equally strong separability of all affective states, as illustrated by the markedly lower MCC obtained for amusement.
\par}

The exploratory FEG-profile descriptors did not consistently improve the most important models. They improved baseline and amusement in the best compact-feature comparison, but they reduced the best stress and meditation results relative to the core interpretable feature set. Therefore, these descriptors should not be treated as the primary physiological explanation. Instead, they are best viewed as a richer nonlinear representation of respiratory variability and predictability that can be evaluated as an extension. The main conclusions can already be supported by simpler and more transparent descriptors: frequency-band energy, FFT band power, wavelet energy/entropy, inspiratory timing, waveform statistics, and Zpm/Zmp autocorrelation transition lags.

This cautious interpretation is supported by the literature balance. Temporal variation in breathing patterns is physiologically meaningful and can be described by autocorrelation-like persistence, nonstationarity, and cycle-to-cycle variability \cite{oku2022temporal}. Recent work also emphasizes that breathing dynamics can modulate emotion and cognition and that entropy or complexity measures may contain affective information \cite{goheen2025breathing_dynamics,soni2019breath_variability}. However, the evidence for autocorrelation transition lags, zero-crossing-derived descriptors, and wavelet-entropy-style markers as validated stress or meditation biomarkers remains weaker than the evidence for rate, phase timing, variability, and spectral/coherence measures. Thus, our Zpm/Zmp and FEG-profile results are best presented as interpretable signal-structure markers and hypothesis-generating features rather than as established physiological biomarkers.

The findings also clarify how this study differs from a purely physiological breathing study. We do not claim that the selected features uniquely measure sympathetic activity, relaxation depth, or meditation quality. Instead, the results show that a small number of respiratory signal descriptors carry subject-independent discriminative information under the WESAD protocol. The added physiological literature provides plausible interpretation for the most stable families of markers, while the LOSO design tests whether those markers are useful across participants.

From a deployment perspective, the raw-signal CNN and the core feature branch have complementary advantages. The 1D-CNN is compact (approximately 0.229M trainable parameters) and uses only a single 3000-sample RESP window, making it a plausible candidate for smartphone, wearable-gateway, or embedded inference after quantization and runtime optimization. This is aligned with recent edge-oriented stress-recognition and TinyML work, where compact models are motivated by limited memory, latency, privacy, and battery constraints \cite{sim2022edge,gibbs2024tinyml,jaiswal2024tinystressnet,cvetkovic2025bioedgenet,abusamah2025tinyml}. The core feature branch is even lighter because it relies on timing, waveform, spectral, wavelet, and autocorrelation-lag summaries that can be computed window by window without multimodal fusion. In contrast, the FEG-Pro extension is computationally heavier and is more appropriate for offline analysis or future optimized implementations. This distinction supports the practical relevance of the stress detector while preserving the interpretability of the feature signatures.

{
The temporal resolution should also be interpreted in relation to the intended application. Safety-critical monitoring of rapidly changing cognitive state can favor substantially shorter analysis windows and fast-response modalities: for example, wearable cognitive-load studies have evaluated 5--30 s windows, with eye-movement features supporting particularly short decision intervals \cite{tervonen2021ultrashort}. By contrast, physiological fatigue and daily-life stress monitoring often operate on longer time scales; a systematic review of driver-fatigue detection reports HRV analyses commonly based on sliding windows of approximately 1--5 min \cite{lu2022driverfatigue}, while a large-scale daily-life stress study used 5 min windows with 4 min overlap, producing one update per minute \cite{smets2018dailystress}. The present 60 s window with a 20 s stride therefore occupies an intermediate regime intended to characterize a sustained respiratory/autonomic state rather than to serve as an emergency-response detector. Predictions are refreshed every 20 s, but if a true state transition occurs inside a 60 s window, the window temporarily contains a mixture of the preceding and new respiratory states. The effective transition latency consequently depends on how rapidly the new physiological pattern dominates the analysis window and is not assumed to equal the stride. Quantifying this transition behavior, and comparing shorter windows such as 20--45 s against the present 60 s context, is an important direction for future real-time evaluation.
\par}

\subsection{Limitations}\label{subsec:limitations}

The study has several limitations that should be considered when interpreting the results.
\begin{itemize}
\item The analysis is based on the WESAD laboratory protocol with 15 subjects. The results therefore require validation on larger cohorts and real-world wearable recordings before clinical or occupational deployment, a limitation repeatedly emphasized in wearable-stress reviews and open-dataset surveys \cite{bolpagni2024personalized,chatzaki2025overview}.
\item {The broader state-specific feature-combination searches remain exploratory signature-discovery analyses rather than fully nested model-selection experiments. However, the specific Top-20 concern for the auxiliary stress search was tested with an additional strict nested analysis: mean MCC decreased from 85.45\% to 82.29\%, indicating modest selection optimism while preserving the main stress-discrimination conclusion. External validation on an independent cohort is still required before treating any selected compact signature as a definitive biomarker set.}
\item {The CNN and feature branches use different internal optimization paths, but they do not differ in the amount of data used for the final outer-fold model fit. The feature models are fitted directly on all 14 non-held-out subjects, whereas the CNN uses an internal TRAIN/VAL split for training-time selection and is then refitted on the same complete 14-subject outer-training pool. The new nested sensitivity analysis addresses the specific auxiliary Top-20 stress search, while the broader state-signature search remains exploratory as noted above; the relevant limitation is therefore model-selection scope rather than a TRAIN+VAL refit asymmetry.}
\item {The 60 s analysis window improves the number of respiratory cycles available for timing, variability, and spectral characterization, but it can blur abrupt state transitions. Although outputs are refreshed every 20 s, transition latency was not measured explicitly in the present WESAD protocol; shorter-window and controlled transition experiments are therefore needed before safety-critical real-time use.}
\item Respiratory-cycle detection passed internal quality control for all analyzed windows, but WESAD does not provide manual breath-cycle annotations. The detector was therefore not externally validated against expert-labelled cycle boundaries.
\item The FEG-Pro descriptors are exploratory and computationally heavier than the core feature families. They should not be treated as established respiratory biomarkers or as the preferred feature set for embedded deployment.
\item Baseline and amusement showed weaker RESP-only separability than stress and meditation. These states may require multimodal physiological, behavioral, or contextual information for robust affect recognition.
\end{itemize}

\section{Conclusion}\label{sec:conclusion}

This study reframes RESP-based affect recognition around state-specific respiratory signatures and compares them with a compact raw-signal CNN. Single-feature threshold screening showed that stress is strongly associated with high-frequency respiratory spectral energy, whereas meditation is strongly associated with breath-cycle aggregated Zmp autocorrelation transition lag. Compact feature-combination analysis further showed that stress is best represented by a continuous balanced-logistic-regression model combining BC-CV Zmp, wavelet energy ratio, skewness, inspiratory time ratio, and signal energy (MCC=83.51), while meditation is best represented by a binary threshold-pattern model combining BC-median Zmp, low-frequency FFT power, wavelet energy entropy, Fre5, and inspiratory timing (MCC=88.65).

The final CNN refit provided the strongest stress detector, achieving accuracy=96.72\%, macro-F1=95.30\%, and MCC=90.61. However, the best compact feature models remained stronger for baseline, amusement, and meditation in terms of MCC. Thus, the practical stress-detection task benefits most from the raw-signal CNN, whereas the feature branch provides stronger and more interpretable state-specific respiratory signatures for several non-stress comparisons.

Adding the full exploratory FEG-profile descriptor set did not consistently improve the most important state signatures. The extension improved baseline and amusement, but it did not improve stress or meditation. Thus, the core physiological interpretation should be based primarily on transparent timing, variability, waveform, spectral/time-frequency, and autocorrelation-lag descriptors. The FEG-profile descriptors remain useful as an exploratory nonlinear extension rather than as the central explanation of respiratory state separability. Overall, the results support a deployment path in which the compact 1D-CNN is used for practical stress detection, while the core feature branch provides transparent respiratory markers for interpretation and lightweight embedded screening.

\section*{Acknowledgements}
The authors thank the creators of the WESAD dataset for making the respiratory recordings available to the research community.

\section*{Declarations}

\paragraph{Funding.}
This research was supported by the Russian Science Foundation (grant no. 22-11-00055-P).

\paragraph{Competing interests.}
The authors declare no competing interests.

\paragraph{Ethics approval and consent to participate.}
This study uses the publicly available, de-identified WESAD dataset and did not involve new data collection from human participants. Ethical approval and informed consent for the original data collection are described in the WESAD dataset publication.

\paragraph{Consent for publication.}
Not applicable.

\paragraph{Data availability.}
The raw dataset analyzed in this study is the publicly available WESAD dataset. Derived window-level features and LOSO summary outputs are available from the corresponding author upon reasonable request.

\paragraph{Code availability.}
The analysis code used for feature extraction, model selection, and model evaluation is available from the corresponding author upon reasonable request.

\paragraph{Materials availability.}
Not applicable.

\paragraph{Author contributions.}
A.V. was responsible for the original research idea, supervision, funding acquisition, methodology, software implementation, formal analysis, and visualization. A.V. and M.T.H. contributed equally to interpretation of results, manuscript writing, critical review, and editing. All authors read and approved the final manuscript.

\begin{appendices}

\newpage
\renewcommand{\thesection}{A}
\renewcommand{\thesubsection}{A.\roman{subsection}}
\refstepcounter{section}
\section*{Appendix A. Basic computation of FEG-Pro forecast-error-growth descriptors}
\label{app:feg_defs}

This appendix summarizes the basic calculation of the exploratory forecast-error-growth descriptors used in this study. The descriptors follow the general FEG-Pro idea of treating forecast-error growth as a structured profile rather than as a single instability number \cite{velichko2026fegpro}. In the present RESP analysis, these variables are used as optional nonlinear features; they are not interpreted as direct clinical respiratory measures.

The input is a scalar respiratory segment
\begin{equation}
    x_1,x_2,\ldots,x_N,
\end{equation}
which is normalized before feature extraction. Local temporal context is represented by sparse history vectors. The delays are chosen from a small autocorrelation-informed lag set, including the Zpm and Zmp transition lags described in Section~\ref{subsec:zpm_zmp}. A generic history vector is
\begin{equation}
    \mathbf{h}_t = \left[x_t, x_{t-d_1}, x_{t-d_2}, \ldots, x_{t-d_m}\right],
\end{equation}
where $d_j$ are selected history delays.

For each history vector $\mathbf{h}_t$, the algorithm finds neighboring histories $\mathbf{h}_q$ and uses their future values to form multi-horizon forecasts. For a horizon $\tau$, a distance-weighted forecast is written schematically as
\begin{equation}
    \widehat{x}_{t+\tau} = \frac{\sum_{q\in \mathcal{N}_k(t)} w_{tq} x_{q+\tau}}{\sum_{q\in \mathcal{N}_k(t)} w_{tq}},
\end{equation}
where $\mathcal{N}_k(t)$ is the set of nearest historical neighbors and $w_{tq}$ decreases with distance in history space. The signed forecast error is
\begin{equation}
    e_t(\tau)=x_{t+\tau}-\widehat{x}_{t+\tau}.
\end{equation}

The error magnitude is summarized over valid anchor times by a geometrically averaged forecast error,
\begin{equation}
    G(\tau)=\exp\left[\frac{1}{M_\tau}\sum_{t=1}^{M_\tau}\log\left(|e_t(\tau)|+\varepsilon\right)\right],
\end{equation}
where $\varepsilon$ prevents numerical singularities for very small errors. The forecast-error-growth curve is then
\begin{equation}
    L(\tau)=\log G(\tau).
\end{equation}
A faster increase of $L(\tau)$ over early horizons indicates a faster finite-horizon loss of predictability.

The main global FEG slope is obtained from a one-line least-squares approximation,
\begin{equation}
    L(\tau) \approx a + \lambda_{\mathrm{FEG}}\tau .
\end{equation}
The corresponding features are the signed slope $\lambda_{\mathrm{FEG}}$ and its absolute value. To capture non-uniform profile geometry, the curve is also summarized using local and curved fits:
\begin{align}
    L(\tau) &\approx a_L + b_L \tau, \quad \tau \leq \tau_c,\\
    L(\tau) &\approx a_R + b_R \tau, \quad \tau > \tau_c,\\
    L(\tau) &\approx a_Q + b_Q\tau + c_Q\tau^2 .
\end{align}
The left and right local slopes describe early- and later-horizon error-growth behavior, while the quadratic approximation gives a mid-horizon slope and a curvature summary. The one-line, two-line, and quadratic fits also produce $R^2$-type fit-quality descriptors, two-line improvement, and curvature gain.

Residual-shape descriptors are calculated after subtracting a smooth fitted trend. If $\widehat{L}_Q(\tau)$ is the quadratic fitted profile, the residual is
\begin{equation}
    r(\tau)=L(\tau)-\widehat{L}_Q(\tau).
\end{equation}
The standard deviation of $r(\tau)$, the roughness of adjacent residual changes, and the second-difference residual variation summarize departures from smooth forecast-error growth. Monotonicity fractions count how often $L(\tau)$ increases or decreases between adjacent horizons. Stability and reliability scores summarize whether estimated slopes remain consistent across fitted profile regions.

Forecast-error distribution entropy (FEDE) descriptors summarize the distributional uncertainty of signed multi-horizon errors or error differences. If $p_j$ are empirical probabilities of a normalized signed-error quantity, the entropy is
\begin{equation}
    H_{\mathrm{FEDE}}=-\sum_j p_j \log(p_j+\varepsilon).
\end{equation}
The FEDE-derived features used in the main text include mean entropy, changes between early and late horizons, and early-horizon FEDE slope. These quantities describe uncertainty and asymmetry in forecast-error distributions rather than only the average rate of error increase.

The FEG-Pro descriptors can be computed once over the whole 60 s respiratory window, over three equal subwindows, or over local breath-cycle segments. When descriptors are computed at the breath-cycle level, the cycle-level values are aggregated using the notation defined in Table~\ref{tab:aggregation}, for example BC-median, BC-IQR, BC-range, or BC-CV. Thus, a feature such as BC-CV \texttt{slope\_two\_left} represents relative breath-to-breath variability of the early local forecast-error-growth slope, whereas BC-median FEDE delta represents a typical cycle-level entropy-change descriptor.

This appendix is deliberately methodological rather than result-oriented. The main analyses first evaluate transparent respiratory descriptors and then add these FEG-Pro descriptors only as an exploratory nonlinear extension. This separation avoids attributing the main physiological conclusions to complex profile descriptors when similar or stronger performance is achieved by simpler timing, spectral, waveform-statistical, and autocorrelation-lag features.

\noindent\textbf{Note.} The formulation above is written specifically for the present RESP feature-extraction pipeline. A detailed standalone presentation and benchmark evaluation of the general FEG-Pro framework is available in \cite{velichko2026fegpro}.

\end{appendices}


\end{document}